\documentclass{llncs}
\usepackage{xypic,epsf,amsmath,amssymb,latexsym,bbm,multicol}
\usepackage[english]{babel}
\usepackage{graphics}
\usepackage{latexsym}
\usepackage{amsmath}
\usepackage{amsfonts}
\usepackage{amssymb}
\usepackage{enumerate}
\newcommand{\limright}{\mbox{\raisebox{2pt}{{\rm lim} $\!\!\!\!\!\!\!\!\!\!\!${\raisebox{-6pt}{$\longrightarrow$}}}}}

\newcommand{\ignore}[1]{}
\newcommand{\Proof}{\noindent {\it Proof\/}:\ }
\newcommand{\QED}{\hspace*{\fill}$\Box$}

\begin{document}
\title{Sheaves and geometric logic and applications to the 
 modular verification of complex systems}

\author{Viorica Sofronie-Stokkermans} 
\institute{Max-Planck Institut {f\"ur} Informatik, Campus E1.4, Saarbr{\"u}cken, Germany \\
Email: {\texttt sofronie@mpi-inf.mpg.de}}

\maketitle

\begin{abstract}
In this paper we show that
states, transitions and behavior of concurrent systems 
can often be modeled as sheaves over a suitable topological space.  
In this context, geometric logic can be used to describe which 
local properties, of individual systems, 
are preserved, at a global level, when interconnecting the systems.
The main area of application is to modular verification 
of complex systems. We illustrate the ideas 
by means of an example involving a family of interacting controllers
for trains on a rail track.
\end{abstract}

\section{Introduction}
\label{intro}

Complex systems, consisting of several components that interact, 
arise in a natural way in a wide range of applications.  
The components may be complex themselves
(they may e.g.\ contain a database; may have 
their specific internal logic and an appropriate inference mechanism; 
a planning mechanism, etc.), or may be simple -  
but even then their composition can complicated because of 
the necessity to take into account the interaction between the 
single components. 
One of the main problems that arise in the verification of such 
complex systems is the state explosion problem:
the state space can grow exponentially with the number of components. 
Symbolic representations of states and symbolic model checking have greatly 
increased the size of the systems that can be verified. 
However, many realistic systems are still too large to be handled.
It is therefore important to find techniques that can be used to 
further extend the size of the systems that can be verified.
One possibility is to check properties in a modular way (i.e.\
verify them for the individual components, infer that they also 
hold in the system obtained by the interconnection of the individual 
components, and then use them to deduce additional properties of 
the system). 
Not all properties are preserved by 
interconnection: for instance deadlocks might occur when interconnecting
deadlock free systems. The main goal of this paper is to offer an answer 
to an important question in verification:
\begin{quote}
{\em Which properties of complex systems can be checked in a modular way? }
\end{quote}
To answer such questions, in this paper we use an analogy with 
phenomena in topology and algebraic geometry,  where sheaves are used to 
describe locally defined objects which can be patched together into a global 
object. Thus, sheaf theory allows to 
establish links between ``local'' and ``global'' properties.
We show that, given a family of interacting systems,  
states, actions, transitions, behavior in time can often be modeled by sheaves 
over a suitable topological space (where the topology expresses how
the interacting systems share the information). 
Many properties of systems can be expressed as assertions about states, 
actions, transitions, behavior in time. The sheaf semantics
allows us to prove, by using 
results from geometric logic, that those 
properties of systems 
that can be expressed by {\em cartesian axioms} 
are preserved after interconnecting the systems. 
 
\medskip
\noindent The starting point of our research is the work of 
Goguen \cite{Goguen92}, who uses sheaves to  
model behavior in an 'interval of observation', and 
Monteiro and Pereira \cite{MonteiroPereira86}, where behavior is 
modeled by sheaves of monoids. 
The idea of modeling 
states, actions and transitions by sheaves with 
respect to a topological space, and of using geometric logic for studying 
the link between properties of the components and properties of the systems 
that arises from their interconnection occurs, to the best of our knowledge, 
for the first time in our previous work  
\cite{Sofronie96a,Sofronie97,Sofronie-Stokkermans99}. 
We present an overview of our results in 
\cite{Sofronie97,Sofronie-Stokkermans99} together with new results 
which illustrate how sheaf 
theory can be used for the modular verification 
of complex systems. We illustrate all the notions introduced by means 
of a running example involving a family of 
interacting controllers controlling a subsets of consecutive trains on 
a linear, loop-free, rail track. 
The main contributions of the paper are summarized below: 
\begin{itemize}
\item We start with a presentation of 
our previous results described in 
\cite{Sofronie96a,Sofronie97,Sofronie-Stokkermans99}, 
where we showed that states, parallel actions, transitions 
and behavior in time can be modeled by sheaves. 
Concerning these topics, the main contribution of this paper
consists in illustrating 
the various notions we use (definition of systems, states, 
parallel actions, transitions, conditions on transition 
relations, categorical constructions, covers, gluing and 
sheaf properties) by means of a running example.
\item In addition to the model of behavior we considered in  
\cite{Sofronie96a,Sofronie97,Sofronie-Stokkermans99}, we also analyze 
a description of behavior by traces of execution (modeled 
by free monoids and partially commutative monoids). We analyze 
gluing and sheaf properties also in this context. 
We pay special attention also in this case to identifying 
situations when the stalks of the sheaves are isomorphic to the behavior 
of the individual systems, whereas the global sections are isomorphic to the 
behavior of the colimit of these systems.
For this, we use results on sheaf representation in universal algebra. 
We establish links with existing results in the study of Petri nets 
and Mazurkiewicz traces \cite{Diekert90} and on modeling behavior 
by sheaves of monoids \cite{MonteiroPereira86}.
\item We use geometric logic for describing 
properties which can be checked modularly. We illustrate the ideas 
on the running example, and describe  a simple complex system 
for trains for which safety and lifeness can be checked in a modular way. 
\end{itemize}

\medskip
\noindent 
{\em Structure of the paper.} The paper is structured as follows.  
In Section~2 we present a model for systems (including also their 
states, parallel actions and transitions). Section~3 contains the definition 
of a category of systems and the description of pullbacks and colimits 
in this category. In Section~4 we give a model for complex, interacting 
systems, and motivate the use of sheaf theory. Sections~5--8 
describe our sheaf-theoretic 
semantics for states, parallel actions, transitions and behavior. 
In Section~9 geometric logic is used to test preservation of 
'local' properties under connection of systems. Several examples 
are given in Section~10.

\section{Systems}

Our aim is to model interconnected systems.  
We assume systems are described by:
\begin{itemize}
\item a set $X$ of control variables of the system, 
a set $\Gamma$ of constraints
on $X$ expressed in a language ${\cal L}$, 
\item a set $A$ of atomic actions, and a set $C$ of constraints on $A$.
\end{itemize}
Let ${\Sigma} = (\mbox{\sf Sort}, O, P)$ be a signature,
consisting of a set {\sf Sort} of sorts, 
a set $O$ of operation symbols and
a set $P$ of predicate symbols. 
For a (many-sorted) set of variables
$X = \{ X_s \}_{s \in \mbox{{\scriptsize {\sf Sort}}}}$
let ${\sf Fma}_{\Sigma}(X)$ be the set of formulae over $\Sigma$. 

\noindent A {\em $\Sigma$-structure} is a structure
$M= ((M_s)_{s \in \mbox{{\scriptsize \sf Sort}}},\{f_M\}_{f \in O},
\{R_M\}_{R \in P})$ where if $f \in O$ has arity 
$s_1 \ldots s_n \rightarrow s$
then $f_M: M_{s_1} \times \ldots \times M_{s_n} \rightarrow M_s$ and
if $R \in P$ has arity $s_1 \ldots s_n$ 
then $R_M \subseteq M_{s_1} \times \ldots \times M_{s_n}$.
The class of all $\Sigma$-structures is denoted ${\sf Str}_{\Sigma}$.
If $M \in {\sf Str}_{\Sigma}$, 
$s : X \rightarrow M$ is a sort-preserving assignment, and 
$\phi \in {\sf Fma}_{\Sigma}(X)$, $(M, s) \models \phi$ 
(abbreviated  by $s \models \phi$) is defined in the 
usual way (cf.\ \cite{ChangKeisler90}, Ch.~1).

\begin{definition} 
A {\em system} $S$ is a tuple $({\Sigma}, X, \Gamma, M, A, C)$, where
\begin{enumerate}
\item[(i)] $\Sigma = ({\sf Sort}, O, P)$  and 
$X = \{ X_s \}_{s \in \mbox{{\scriptsize {\sf Sort}}}}$ are as specified above;
together they define the
{\em language ${\cal L}_S$ of the system $S$};
\item[(ii)] $\Gamma \subseteq $ {\sf Fma}$_{\Sigma}(X)$ 
is a set of constraints,
which is closed with respect to the semantical consequence 
relation\footnote{The
relation $\models_M$ is defined by
$\Gamma \models_M \phi$ if and only if for every assignment 
$s : X \rightarrow M$ of
values in $M$ to the variables in $X$,
if $s \models \gamma$ for every $\gamma \in \Gamma$,
then $s \models \phi$.} $\models_M$;
\item[(iii)] $M \in {\sf Str}_{\Sigma}$;
\item[(iv)] $A$ is a set of actions; for every $a \in A$, a set
$X^a \subseteq X$ of variables on which $a$ depends,
and a transition relation $Tr^a \subseteq St^a \times St^a$, where
$St^a = \{ s_{|X^a} \mid s : X \rightarrow M,  s \models \Gamma \}$ are 
specified;
\item[(v)] $C$ is a set of constraints on actions, 
expressed by boolean equations
over $F_B(A)$ (the  free boolean algebra generated
by $A$) stating e.g.\ which actions can (or have to) be executed in parallel,
and which cannot; $C$ must contain all boolean equations that can 
be deduced from $C$.
\end{enumerate}
\end{definition}

\noindent 
In what follows, we may refer to any of the components of a system $S$ by
adding $S$ as a subscript, e.g.\ $\Sigma_S$ for its signature.
$X_S^a$ will denote the minimal set of variables on which $a
\in A_S$ depends, and $Tr_{S}^a$ the transition 
relation associated with $a$. 

\noindent  For the sake of simplicity, in the examples below 
we will only mention explicitly the axioms in $\Gamma$ and $C$ and not all 
their consequences.

\begin{example}
\label{example-system}
We consider a system consisting of $n$ consecutive trains on a linear track 
controlled by a radio controller 
(cf.\ also \cite{Jacobs-Sofronie-pdpar06}). 
The trains report their position to the controller at fixed time 
intervals $\Delta t$. 
The controller analyzes the distances between successive trains 
(we assume that certain security distance treshholds 
$l_0 <  l_1 < \dots < l_m < \dots$ 
and corresponding maximal speed limits 
${\sf maxSpeed}(1) < \dots < {\sf maxSpeed}(m) < \dots$, 
deemed to be safe for the trains, are known) and  
updates the movement modes of trains accordingly. 
A train with movement mode $k$ can  move in the next time interval $\Delta t$ 
with an arbitrary speed between a minimal speed and 
the maximal speed limit of mode $k$, ${\sf maxSpeed}(k)$. 

The system is modeled as follows:

\begin{enumerate}
\item[(i)] {\em Language:} $\Sigma = ({\sf Sort}, O, P)$, 
where ${\sf Sort} = \{ {\sf real}, {\sf nat} \}$; 
\begin{itemize}
\item $O = \{ +,-,    
{\sf minSpeed}, {\sf maxSpeed}, {\sf succ} \}$, 
where: 
\begin{itemize}
\item $+, -$ are function of arity ${\sf real,real} {\rightarrow} {\sf real}$, 
\item ${\sf minSpeed}$ is a constant of sort ${\sf real}$,
\item ${\sf maxSpeed}$ a function of arity 
${\sf nat} \rightarrow {\sf real}$, and 
\item ${\sf succ}$ of arity ${\sf nat} \rightarrow {\sf nat}$. 
\end{itemize}
\item $P = \{\leq \}$, where 
$\leq$ has arity ${\sf real}, {\sf real}$. 

\item $X = \bigcup_{i = 1}^n \{ {\sf TrainIndex}_i, {\sf ActualPos}_i, 
{\sf RepPos}_i, {\sf Mode}_i \}$, 
where ${\sf TrainIndex}_i$ controls the number of train $i$ on the line track, 
and ${\sf ActualPos}_i, {\sf RepPos}_i$ and ${\sf Mode}_i$ 
control the actual, resp.\ reported position and the 
movement mode of train $i$ respectively.
\end{itemize} 
\medskip
\item[(ii)] {\em Constraints:} 
$\Gamma = \{ {\sf succ}({\sf TrainIndex}_i) = 
{\sf TrainIndex}_{i+1} \mid i \in \{ 1, \dots, n-1 \} \}$. 
\medskip
\item[(iii)] {\em Model} $M {=} (M_{\sf nat},M_{\sf real},
+, -,{\sf minSpeed},{\sf maxSpeed}, {\sf succ}, \leq)$, where:
\begin{itemize}
\item The universes are: 
\begin{itemize}
\item $M_{\sf nat} = {\mathbb N}$; $M_{\sf real} = {\mathbb R}$; 
\end{itemize}
\item The operations are defined as follows:
\begin{itemize}
\item $+, -$ are addition and subtraction on ${\mathbb R}$,  
\item ${\sf succ} : {\mathbb N} \rightarrow {\mathbb N}$ 
is the successor function, 
\item ${\sf minSpeed} \in {\mathbb R}$,
\item ${\sf maxSpeed} : {\mathbb N} \rightarrow {\mathbb R}$ 
associates with a mode $k \in {\mathbb N}$ 
the maximal allowed speed in mode $k$; 
\end{itemize}
\item The predicates are defined as follows:
\begin{itemize}
\item $\leq$ is the order relation on ${\mathbb R}$. 
\end{itemize}
\end{itemize}
\medskip
\item[(iv)] {\em Actions:} 
$A = \{ {\sf report}_i \mid i \in \{ 1, \dots, n \} \} \cup 
\{ {\sf update} \} \cup  \{ {\sf move}_i \mid i \in \{ 1, \dots, n \} \}$.

\noindent 
\begin{itemize}
\item ${\sf report}_i$ depends on the variables 
$X^{r_i} = \{ {\sf ActualPos}_i, {\sf RepPos}_i, {\sf Mode}_i \}$. 

\noindent 
If $s, s' : X \rightarrow M$ then 
$(s_{|X^{r_i}}, s'_{|X^{r_i}}) \in Tr^{r_i}$ 
iff the following hold: 
\begin{itemize}
\item $s({\sf Mode}_i) = 0$ 
\item $s'({\sf RepPos}_i) = s({\sf ActualPos}_i)$ 
\item $s'({\sf ActualPos}_i) = s({\sf ActualPos}_i).$ 
\end{itemize}

\item ${\sf update}$ depends on  $X^u = \bigcup_{i \in \{ 1, \dots, n \}} 
\{ {\sf ActualPos}_i, {\sf RepPos}_i, {\sf Mode}_i \}$.  

\noindent If $s, s' : X \rightarrow M$ then 
$(s_{|X^{u}}, s'_{|X^{u}}) \in Tr^{r_i}$ iff  
for all $i \in \{ 1, \dots, n \}$ the following hold: 
\begin{itemize}
\item $s({\sf Mode}_i) = 0$, 
\item $s'({\sf ActualPos}_i) = s({\sf ActualPos}_i)$, 
\item $s'({\sf RepPos}_i) = s({\sf ActualPos}_i)$, and 
\item $s'({\sf Mode}_i)$ is updated according to the following rules: 
$s'({\sf Mode}_1) > 0$ and for all $i \geq 2$:
\end{itemize}

\smallskip
$~~~~~$ if $l_k < s({\sf RepPos}_{i-1}) - s({\sf RepPos}_{i}) \leq l_{k+1}$ then $s'({\sf Mode}_{i}) = k+1$. 

\smallskip
\item ${\sf move}_i$ depends on 
$X^{m_i} = \{ {\sf ActualPos}_i, {\sf Mode}_i \}$. 

It is 
enabled at a state $s$ iff $s({\sf Mode}_i) > 0$ for all 
$i \in \{ 1, \dots, n \}$; it 
changes ${\sf ActualPos}_i$ according to the value of ${\sf Mode}_i$
as follows, for $i \in \{ 1, \dots, n \}$: 
 $s'({\sf ActualPos}_i) {\in} [{\sf PosMin}, {\sf PosMax}]$, where: 
 
\begin{itemize}
\item ${\sf PosMin} = {\sf RepPos}_i {+} \Delta t {*} {\sf minSpeed}$, 
\item ${\sf PosMax} = {\sf RepPos}_i {+} \Delta t {*} {\sf maxSpeed}(s({\sf Mode}_i))$;
\end{itemize} 

\smallskip 
\noindent 
and it updates the value of ${\sf Mode}_i$ to 0: 
$s'({\sf Mode}_i) = 0$ for $i \in \{ 1, \dots, n \}$.
\end{itemize}
\medskip
\item[(v)] {\em Constraints on actions:} 
$ C = \{ {\sf report}_1 = {\sf report}_2 = \dots = 
{\sf report}_n = {\sf update} \} \cup \\ 
\{ {\sf report}_i \wedge {\sf move}_i = 0 \mid i \in \{ 1, \dots, n \} \}
\cup \{  {\sf move}_1 = \dots =  {\sf move}_n \}.$ 
\end{enumerate}
\end{example}

\subsection{States, parallel actions}
It is important to describe {\em the states} of a system and  
{\em the actions which can be performed in parallel} (which we here 
name admissible parallel actions).

\begin{definition} 
Let $S = (\Sigma, X, \Gamma, M, A, C)$ be a system. 
\begin{itemize}
\item A {\em state} of $S$ is an assignment $s : X \rightarrow M$
satisfying all formulae in $\Gamma$.
The  {\em set of states} of the system $S$ is 
$St(S) = \{ s : X \rightarrow M \mid s \models \Gamma \}$.
\item The {\em admissible parallel actions} of 
$S$ are sets of actions, represented by maps
$f : A  \rightarrow \{ 0, 1 \}$ that satisfy all constraints in $C$.
The {\em set of admissible parallel actions} of $S$ is the set 
$Pa(S) = \{f : A  \rightarrow \{ 0, 1 \} \mid f \mbox{ satisfies } C \}$.
\end{itemize}
\end{definition}

\noindent Below we restrict our attention to {\em finite} systems, 
i.e.\ systems whose signatures, sets of
control variables and sets of actions are finite; this 
suffices for practical applications and avoids having to 
consider infinitely many actions occurring in parallel.
\begin{example}
Consider the system $S$ in Example~\ref{example-system} with $n \geq 2$. 
A state is a map $s : X \rightarrow M$ which satisfies $\Gamma$. 
For instance, any map $s : X \rightarrow M$ such that:  
\begin{itemize}
\item $s({\sf TrainIndex}_1) = 1, s({\sf TrainIndex}_2) = 2, \dots, 
s({\sf TrainIndex}_n) = n$ or 
\item  $s({\sf TrainIndex}_1) = 100, s({\sf TrainIndex}_2) = 101, {\dots}, 
s({\sf TrainIndex}_n) = 100{+}(n{-}1)$. 
\end{itemize}
is a state of $S$. If $s({\sf TrainIndex}_1) = 1$ and 
$s({\sf TrainIndex}_2) = 3$, $s$ cannot be a state.

\smallskip
\noindent An admissible parallel action is a map 
$f : A \rightarrow \{ 0, 1 \}$ which 
satisfies the constraints in $C$. 
Examples of admissible parallel actions 
are 
\begin{enumerate}
\item $f({\sf report}_1) \,{=}\, f({\sf report}_2) \,{=}\, \dots \,{=}\, 
f({\sf report}_n) \,{=}\, f({\sf update}) \,{=}\, 1$,  and $0$ otherwise, 
\item $f({\sf move}_1) = \dots = f({\sf move}_n) = 1$ and $0$ otherwise.
\end{enumerate}
Any map $f$ with
$f({\sf move}_1) = f({\sf report}_1) = 1$, or 
with $f({\sf report}_i) = 0$ but $f({\sf update}) = 1$, 
is not an admissible 
parallel action, since it does not satisfy the constraints in $C$. 
\end{example}

\subsection{Transitions} 
Let $S = (\Sigma, X, \Gamma, M, A, C)$ be a system. 
Let $Tr_S(a) = \{ (s_1, s_2) \mid s_1, s_2 \in St(S), 
({s_1}_{|X^a},{s_2}_{|X^a}) \in Tr^a, s_1(x) = s_2(x) 
\mbox{ if } x \not\in X^a \}$. 
We extend the notion of transition to parallel actions. 
For this we present two (non-equivalent) properties of transitions that express
compatibility of the actions in an admissible parallel action: 

\smallskip
\noindent \begin{tabular}{@{}ll}
({\bf Disj}) & Let $f \in Pa(S), s \in St(S)$ such that
for every $a \in f^{-1}(1)$ there is an $s^a \in$ \\
& $St(S)$ with  $(s_{|X^a}, s^a_{|X^a}) \in Tr^a$. Then 
for all $a,b \in f^{-1}(1)$ and $x {\in} X^{a} {\cap} X^{b}$, \\
& $s^{a}(x) = s^{b}(x)$ (the new local states
agree on intersections). 
Then, 
\end{tabular}

\noindent $\begin{array}{@{}l@{}l}
~~~~~~~~~~~~~Tr_S(f)  =  \{(s,t) \mid~& s, t \in St(S),
({s}_{|X^a},{t}_{|X^a}) \in Tr^a \mbox{ for every } a
\mbox{ such that }  \\
& f(a) = 1 \mbox{ and }
s(x) = t(x) \mbox{ if }
x \not\in \bigcup_{a, f(a) = 1}X^a \}.
\end{array}$

\medskip
\noindent The property ({\bf Disj}) applies when
a parallel action $f : A \rightarrow \{ 0, 1 \}$ is admissible
iff its components do not consume common resources.
This happens e.g.\ if
for all $a_1, a_2 \in A$ with $f(a_1) = f(a_2) = 1$, either $a_1 = a_2 \in C$ 
or $X^{a_1}$ and $X^{a_2}$ are disjoint. 
In concurrency theory, this property is called  ``real parallelism'' or 
``independence''.

\begin{example}
\label{example-disj}
Consider the example in Section~\ref{example-system}. 
Let $f : A \rightarrow \{0, 1 \}$ be an admissible parallel action.
We have two possibilities: 

\begin{enumerate}
\item[(i)] $f({\sf report}_1) = \dots = f({\sf report}_n) = 
f({\sf update}) = 1$ and $0$ otherwise. 

\noindent 
The transition relation of this parallel action updates the value of each 
variable ${\sf RepPos}_i$ according to the transition relation of 
${\sf report}_i$, resp.\ ${\sf update}$. The changes are not 
contradictory, since the effect of ${\sf update}$ agrees with the effect of 
${\sf report}_1, \dots, {\sf report}_n$ on the variables in 
$X^u \cap X^{r_i}$. Thus, ({\bf Disj}) holds. 
 
\medskip
\item[(ii)] 
$f({\sf report}_1) = \dots = f({\sf report}_n) = f({\sf update}) = 0$ and 
$f({\sf move}_1) = \dots = f({\sf move}_n) = 1$ 
and $f$ is 0 otherwise. As the actions ${\sf move}_j, j = 1, \dots, n$ 
depend on disjoint sets of variables, ({\bf Disj}) 
is satisfied also in this case.

\noindent 
The transition relation of this parallel action updates the value of each 
variable ${\sf ActualPos}_i$. 
Since the sets of variables these actions depend upon, namely  
$X^{m_i}$, are mutually disjoint, these changes cannot be  
contradictory.
\end{enumerate}
\end{example}

\noindent \begin{tabular}{@{}rl}
({\bf Indep}) & 
Assume that if $a = b \in C$ then $X^a = X^b$ and $Tr^a = Tr^b$, 
and $a$ and $b$ \\
& can both be identified with one action: the parallel execution of 
$a$, $b$.\\
& Let $f \in Pa(S), s \in St(S)$.  We identify all $a, b \in A$ with 
$a = b \in C$ \\
& and $f(a) = f(b) = 1$. Let  $\{b_1, \dots, b_m \} \subseteq f^{-1}(1)$.
We assume that: \\[1ex]
~~~~~(i) & $g : A \rightarrow \{ 0, 1 \}$, defined by $g(a) = 1$ iff 
$a \in \{b_1, \dots, b_m \}$, is in $Pa(S)$;  \\[1ex]
~~~~(ii) & if $s \stackrel{b_1}{\longrightarrow}  
s_1 \stackrel{b_2}{\longrightarrow} 
s_2 \longrightarrow \dots \longrightarrow s_{m-1} \stackrel{b_m}{\longrightarrow} t$ 
then for every permutation \\
& $\sigma$ of $\{ 1, \dots, m \}$, there exist states 
$t_1^{\sigma}, t_2^{\sigma},\dots, t_{m-1}^{\sigma}$ such that we have 
\end{tabular}
$$s \stackrel{b_{\sigma(1)}}{\longrightarrow} t_1^{\sigma} 
\stackrel{b_{\sigma(2)}}{\longrightarrow} t_2^{\sigma} \longrightarrow \dots
\longrightarrow t_{m-1}^{\sigma} 
\stackrel{b_{\sigma(m)}}{\longrightarrow} t$$ 
\bigskip
\noindent 
In this case we define a the transition associated with a parallel action $f$ by: 
$$\begin{array}{@{}l@{}l} 
Tr_S(f)  = \{(s,t) \mid~ & s, t \in St(S), \mbox{ and } \exists 
s_0, s_1, \dots, s_{n-1}, s_n \in St(S) \\
 & \mbox{such that }  s_0 = s \text{ and } s_n = t, \text{ and } \mbox{for all } i \text{ with }\\
 & 1 \leq i \leq n, ~s_{i-1}, s_i) \in Tr_S(a_i)   \}.
\end{array}$$
It is easy to see that if $(s,t) \in Tr_S(f)$ then
$s(x)  =  t(x)$ for every $x \not\in \bigcup_{a, f(a) = 1}X^a$.

\bigskip
\noindent 
The property ({\bf Indep}) reflects how transitions are interpreted  
when actions to be performed in parallel do consume common resources.
It applies if the state reached after executing an action
is uniquely determined: 
the fact that all components of a parallel action
$f : A \rightarrow \{ 0, 1 \}$ can be applied at a state $s$
is a necessary
condition for $f$ to be applicable at state $s$,
but in general not sufficient (in addition, one has to ensure that there are
enough resources to perform all actions).
Condition ({\bf Indep})(i) holds e.g.\ if $C$ is the set of all 
consequences of a set $C_0$ consisting only of formulae of the form 
$a_1 = a_2$ and $a_1 \wedge a_2 = 0$. 
Condition ({\bf Indep})(ii) states that the final state does not depend on 
the order in which the actions 
are executed 
(it is related to the notions of interleaving and permutable actions 
used in concurrency).

\begin{example}
\label{example-indep}
We consider a variant of Example~\ref{example-system}, in which we assume 
that there is no control unit, but all trains have access to 
all information about the positions of all trains.   The trains report all 
together and move all together. 
The actions are 
$A = \{ {\sf report}_1, \dots, {\sf report}_n \} \cup 
\{ {\sf move}_1, \dots, {\sf move}_n \}$, with constraints 
$C = \{ {\sf report}_1 = \dots = {\sf report}_n \} \cup 
\{ {\sf move}_1 =  \dots =  {\sf move}_n \} \cup 
\{ {\sf report}_i \wedge {\sf move}_i = 0 \mid i \in \{ 1, \dots, n \} \}$. 

\noindent Let $f : A \rightarrow \{ 0, 1 \}$ be an admissible parallel action.
Then  $f^{-1}(1)$ is either $\emptyset$ or 
$\{ {\sf report}_1, \dots, {\sf report}_n \}$ or 
$\{ {\sf move}_1, \dots, {\sf move}_n \}$. 
As in all cases the actions in $f^{-1}(1)$ depend on disjoint sets of 
variables, the final state does not depend on the order in which the actions 
would be performed sequentially.
\end{example}

\section{A category of systems}
Essential to our model for communication 
is that systems have common subsystems through 
which information exchange is made.
Let $S, T$ be two systems.
We say that $S$ is a {\em subsystem} of 
$T$ (denoted $S \succ\!\rightarrow T$) if
${\Sigma}_{S} \subseteq {\Sigma}_{T}$,
$X_S \subseteq X_T$,
$A_S  \subseteq  A_T$, the
constraints in $\Gamma_S$ (resp.\ $ C_S$)
are consequences of the
constraints in $\Gamma_T$ (resp.\ $C_T$),
and $M_S = {M_T}_{|\Sigma_S}$
(the reduct of $M_T$ to the signature $\Sigma_S$). 

\smallskip
\noindent Let $S \succ\!\rightarrow T$.  If we regard a transition in $T$ 
from the perspective of $S$, 
some variables in $S$ may change their values with 
no apparent cause, namely 
if some action in $A_T$ but not in $A_S$ is performed, which
depends on variables in $X_S$.
If this cannot be the case, we call the subsystem $S \succ\!\rightarrow T$
{\em transition-connected}.
Formally: 

\begin{definition}
$S$ is a {\em transition-connected (t.c.)
subsystem of} $T$ (denoted $S \hookrightarrow T$)
if $S \succ\!\rightarrow T$
and the following two conditions hold:
\begin{description}
\item[($T_1$)] If $a \in A_T$ and $X_T^a \cap X_S \neq \emptyset$ then
$a \in A_S$, and  $X^a_S = X^a_T \cap X_S$. 
\item[($T_2$)] If $a \in A_S$, $s_1, s_2 \in St(T)$, and 
$({s_1}_{|X_T^a}, {s_2}_{|X_T^a}) \in Tr_T^a$ then
$({s_1}_{|X_S^a}, {s_2}_{|X_S^a}) \in Tr_S^a$.
\end{description}
\end{definition}
\noindent It is easy to see that the relation $\hookrightarrow$ 
is a partial order on systems.

\begin{example}
\label{example-transition-connected}
Consider the system $S = (\Sigma, X, \Gamma, M, A, C)$ 
in Example~\ref{example-system}. Let $k$ and $l$ be such that $1 \leq k \leq l \leq n$ and let 
$I = \{ k, \dots, l \}$. Consider the restriction 
$S_k^l = (\Sigma, X_k^l, \Gamma_k^l, M, A_k^l, C_k^l)$ of $S$ to 
the consecutive trains controlled by the variables in 
$\{ {\sf TrainIndex}_i \mid i \in I \}$. 

\begin{itemize}
\item $X_k^l = \bigcup_{i \in I} \{ {\sf TrainIndex}_i, 
{\sf ActualPos}_i, {\sf RepPos}_i, 
{\sf Mode}_i \}$, 
\item $\Gamma_k^l = \{ {\sf succ}({\sf TrainIndex}_i) = 
{\sf TrainIndex}_{i+1} \mid i \in \{ k, \dots, l-1 \} \}$, 
\item $A_k^l = \{ {\sf report}_i \mid i \in I \} \cup \{ {\sf update } \} \cup 
\{ {\sf move}_i \mid i \in I \}$, and 
\item $C_k^l$ is the restriction of $C$ to the actions in $A_k^l$:

\noindent $
\begin{array}{@{}ll}
C_k^l = & \{ {\sf report}_i = {\sf update} \mid i \in I \} \cup 
\{ {\sf report}_i \wedge {\sf move}_i = 0 \mid i \in I \} \cup \\ 
& \{ {\sf move}_k = \dots = {\sf move}_l \}.
\end{array}$ 
\end{itemize}
Condition $(T_1)$ obviously holds: 
if an action of $S$ 
depends on variables known in $S_k^l$, 
then the action is known in $S_k^l$.
Condition $(T_2)$ obviously holds for 
$\{ {\sf report}_i \mid i \in I  \}  \cup 
\{ {\sf move}_i \mid i \in I \}$ and, for  
${\sf update}$,  for all 
trains which follow a train known in $S_k^l$.
For the first train 
$(T_2)$ is a consequence of the fact that
the mode update restrictions in $S$ are stronger than those in $S_k^l$
(any mode allowed in $S$ is still allowed in $S_k^l$). 
\end{example}

We define a category {\sf TcSys} having as objects systems, and a morphism
$S \hookrightarrow T$ between $S$ and $T$ whenever $S$ is a t.c.\ 
subsystem of $T$.
{\sf TcSys} has {\em pullbacks} (infimums with respect to this order 
of t.c.\ subsystems of a given system; 
we will denote this operation by $\wedge$) and 
{\em colimits} of diagrams of t.c.\ subsystems of a given system.

\begin{proposition}
The category {\sf TcSys} has pullbacks.
\end{proposition}
\Proof  Let $S_1 \hookrightarrow S$ and $S_2 \hookrightarrow S$,  
where $S = (\Sigma, X, \Gamma, M, A, C)$, 
$S_i = (\Sigma_i, X_i, \Gamma_i,$ $M_i, A_i, C_i)$. 
Then $M_i = M_{|\Sigma_i}$, 
 and for every 
$a \in A_i$, $X_i^a = X_S^a \cap X_i$ (i = 1, 2)
Hence, for every $a \in A_1 \cap A_2$,
$X_1^a \cap X_2 = X_2^a \cap X_1 =  X_S^a \cap X_1 \cap X_2$. 

Let $S_{12} = ( \Sigma_1 {\cap} \Sigma_2, X_1 {\cap} X_2, 
\Gamma_1 {\cap} \Gamma_2, {M_S}_{|\Sigma_1 \cap \Sigma_2},  A_1 {\cap} A_2,
 C_1 {\cap} C_2)$, and such that for every 
$a \in A_1 {\cap} A_2$, 
$X_{12}^a = X_1^a {\cap} X_2 = X_2^a {\cap} X_1 = 
 X_S^a {\cap} X_1 {\cap} X_2$, and 
$Tr_{12}^a = \{ ({s_1}_{|X_{12}^a}, {s_2}_{|X_{12}^a}) \mid 
s_1, s_2 \in St(S_1), ({s_1}_{|X_1^a},{s_2}_{|X_1^a}) \in Tr_{S_1}^a \} 
\cup \{ ({s_1}_{|X_{12}^a}, {s_2}_{|X_{12}^a}) \mid 
s_1, s_2 \in St(S_2), ({s_1}_{|X_2^a},{s_2}_{|X_2^a}) \in Tr_{S_2}^a \}$.
It is easy to see that $S_{12}$ is a transition-connected subsystem 
of both $S_1$ and $S_2$, and has the universality property of 
a pullback. \QED

\begin{proposition}
Let $S = (\Sigma, X, M, \Gamma, A, C)$ be a system and 
$\{ S_i \hookrightarrow S \mid i \in I \}$
a family of transition-connected subsystems of $S$, 
where for every $i \in I$, 
$S_i = (\Sigma_i, X_i, M_i, \Gamma_i, A_i, C_i)$.
The colimit of this family in ${\sf SYS_{il}}$ is the system
${\overline S}$ with:
\begin{itemize}
\item $\Sigma_{\overline S} = \bigcup_{i \in I} \Sigma_i$, 
\item $X_{\overline S} = \bigcup_{i \in I} X_i,$ 
\item $M_{\overline S} =  M_{|\bigcup_{i \in I} \Sigma_i},$  
\item $\Gamma_{\overline S} = (\bigcup_{i \in I} \Gamma_i)^{\bullet}$
(the family of all logical consequences of $\bigcup_{i \in I} \Gamma_i$),    
\item $A_{\overline S} = \bigcup_{i \in I} A_i,$ 
\item $C_{\overline S} = (\bigcup_{i \in I} C_i)^{\bullet}$
(the family of all logical consequences of $\bigcup_{i \in I} C_i$), 
\end{itemize}
and where for every $a \in  \bigcup_{i \in I} A_i$  
$X_{\overline S}^a = \bigcup_{a \in A_{i}} X_i^a$, and 
$Tr_{\overline S}^a =
 \{ ({s_1}_{|X_{\overline S}^a},{s_2}_{|X_{\overline S}^a}) \mid  
s_1, s_2 \in St({\overline S}), 
\mbox{ and for every }
i \in I \mbox{ with } a \in A_i, 
({s_1}_{|X_i^a},{s_2}_{|X_i^a}) \in Tr_{S_i}^a \}$.
\end{proposition}
\Proof (Sketch) One needs to show that for every $i \in I$, 
$S_i$ is a transition-connected subsystem 
of ${\overline S}$, and that 
${\overline S}$ satisfies the universality property 
of a colimit. The proof is long, but straightforward. \QED

\begin{example}
Consider the system $S$ in Example~\ref{example-system}, 
and two restrictions $S_1 = S_k^n$ and $S_2 = S_1^l$ 
constructed as in 
Example~\ref{example-transition-connected}. 
The pullback of $S_1$ and $S_2$ is 
$S_{12} = S_k^l$ (defined as 
in Example~\ref{example-transition-connected} if $k \leq l$, or 
the system with the empty set of control variables 
and actions if $l < k$).
The colimit ${\overline S}$ of the diagram 
$\{ S_1, S_2, S_{12} \}$ 
(with transition-connected morphisms $S_k^l \hookrightarrow S_k^n, 
S_k^l \hookrightarrow S_1^l$ has the following components: 
\begin{itemize}
\item $\Sigma_{\overline S} = \Sigma; M_{\overline S} = M$; 
\item $X_{\overline S} = 
\bigcup_{i \in \{ 1, \dots, l \} \cup \{ k, \dots, n \}} 
\{ {\sf TrainIndex}_i, {\sf ActualPos}_i, {\sf RepPos}_i, 
{\sf Mode}_i \}$,
\item $\Gamma_{\overline S} = \{ {\sf succ}({\sf TrainIndex}_i) = 
{\sf TrainIndex}_{i+1} \mid i \in \{ 1, \dots, l-1 \} \cup 
\{k, \dots, n-1 \} \}^{\bullet}$,
\item $A_{\overline S} = 
\bigcup_{i \in \{ 1, \dots, l\} \cup \{ k, \dots, n \}} 
\{ {\sf report}_i, {\sf move}_i \} \cup \{ {\sf update } \}$; 
\item $C_{\overline S} = (\{ {\sf report_i} = 
{\sf update} \mid i \in \{ 1, \dots, l \} \cup \{ k, \dots, n \} \} \cup$ \\ 
$\mbox{\hspace{1cm}} \{ {\sf report}_i \wedge {\sf move}_i = 0 \mid  i \in 
\{ 1, \dots, l \} \cup \{ k, \dots, n \}\} \cup$ \\
$\mbox{\hspace{1cm}} \{ {\sf move}_1 = \dots = {\sf move}_l \} \cup \{ {\sf move}_k = \dots = {\sf move}_n \})^{\bullet}$.
\end{itemize}
If $k \leq l$ then ${\overline S}$ coincides with $S$.
If $l < k-1$ then $X_{\overline S} \neq X$, so ${\overline S}$ is 
obviously different from $S$. 
Assume now that $l = k-1$. Then $X_{\overline S} = X, A_{\overline S} = A, 
C_{\overline S} = C$, but $\Gamma_{\overline S} \neq \Gamma$
(the constraint 
${\sf succ}({\sf TrainIndex}_{k-1}) = {\sf TrainIndex}_k$ 
cannot be recovered from $\Gamma_1^l \cup \Gamma_k^n$), hence 
${\overline S}$ is different from $S$ also in this case. 
\label{ex-colimit}
\end{example}

\section{Modeling families of interacting systems}
\label{insys}
When analyzing concrete complex systems, 
we tend to be interested in a
subcategory of {\sf TcSys}, containing only the systems relevant
for a given application. 
To this end, we assume a family {\sf InSys} of interacting systems
is specified, fulfilling:   
\begin{enumerate}
\item All $S \in {\sf InSys}$ are transition-connected subsystems of a 
system ${\overline S}$ with $A_{\overline S}$ finite.
\item {\sf InSys} is closed under all pullbacks $S_1 \wedge S_2$ of t.c.\ subsystems $S_1, S_2$
of ${\overline S}$. 
\item $({\sf InSys}, \wedge)$ is a meet-semilattice.
\end{enumerate}
The first condition enforces the compatibility of models on common sorts
and the finiteness of $A_S$ for every $S \in {\sf InSys}$; 
the second and third condition ensure that all systems by which 
communication is handled are taken into account.
A system obtained by interconnecting some elements
of {\sf InSys} can either be seen 
as the set of all elements of {\sf InSys} by whose interaction 
it arises (a subset of {\sf InSys} which is downwards-closed with respect to
$\hookrightarrow$) or as the colimit of 
such a family of elements.
We define $\Omega({\sf InSys})$ as consisting of all families of elements of
{\sf InSys} which are closed under transition connected subsystems. Clearly,
$\Omega({\sf InSys})$ is a 
topology on ${\sf InSys}$.

\smallskip
\noindent {\bf Note:} 
It is easy to see that $\Omega({\sf InSys})$ 
is the Alexandroff topology associated with the 
dual of the poset $({\sf InSys}, \hookrightarrow)$. 
Since we assumed that ${\sf InSys}$ is finite and closed under pullbacks, 
this topology coincides with the Scott topology associated with the 
dual of $({\sf InSys}, \hookrightarrow)$. 

\begin{example}
\label{interconnect-3sys}
Consider now the extension of the example in 
Section~\ref{example-system} considered in Example~\ref{ex-colimit}: 
Let $k \leq l \in \{ 1, \dots, n \}$, 
let $I_1 = \{ k, \dots, n \}, I_2 = \{ 1, \dots, l \}, 
I_{12} = \{ k, \dots, l \}$, and 
let ${\sf InSys} = \{ S_1, S_2, S_{12}\}$ 
be the family consisting of the subsystems of 
$S = (\Sigma, X, \Gamma, M, A, C)$ described in Section~\ref{example-system}
corresponding to the sets of trains with indices in $I_1, I_2$ and $I_{12}$ 
respectively: $S_1 = S_k^n$, $S_2 = S_1^l, S_{12} = S_k^l$.
Then ${\sf InSys}$ satisfies conditions (i), (ii) and (iii) above.
The system obtained by interconnecting $S_1, S_2, S_{12}$ 
can be regarded either as the set $\{ S_1, S_2, S_{12} \}$ or as the colimit 
of the diagram defined by these systems, which coincides with 
the system $S$ defined in Section~\ref{example-system}. 
In this 
case, $\Omega({\sf InSys})$ consists of the following sets 
$\{ \emptyset, \{ S_{12} \}, \{ S_1, S_{12} \}, 
\{ S_2, S_{12} \}, \{ S_1, S_2, S_{12} \} \}$.
\end{example}

\noindent Our goal is to express the links between components of a system and 
the result of their interconnection. 
We start from the observation that compatible local states 
can be 'glued' into a global state 
(similar for parallel actions, transitions). 
For expressing 
such gluing condition in a general setting, we use sheaf theory.

\subsection{Sheaf theory: An introduction}
\label{sheaves}
In what follows, notions from category theory
are assumed to be known. For definitions and details  we refer to \cite{Johnstone82} or 
\cite{MacLaneMoerdijk92}. (In what follows categories and sheaves
will be denoted in sans-serif style, e.g.\ {\sf Set}, {\sf Sh}$(I)$.)

\smallskip
\noindent Let $I$ be a topological space, and $\Omega(I)$ the topology on $I$.

\begin{definition} 
A {\em presheaf}
on $I$ is a functor $P : \Omega(I)^{op} \rightarrow $ {\sf Sets}.
Let $U \subseteq V$ be open sets in $I$, and 
$i_U^V : U \hookrightarrow V$ the inclusion morphism in
$\Omega(I)$.  The restriction to $U$, 
$P(i^V_U): P(V) \rightarrow P(U)$ is denoted by $\rho^V_U$.

\noindent A {\em sheaf} on $I$ is
a presheaf $F : \Omega(I)^{op} \rightarrow $ {\sf Sets} that
satisfies the following condition: 

\begin{quote}
\begin{enumerate}
\item[]for each open cover $(U_i)_{i \in I}$ of $U$ and family of elements
$s_i {\in} F(U_i)$ s.t.\ for all $i,j$, $\rho^{U_i}_{U_i\cap U_j}(s_i)
{=} \rho^{U_j}_{U_i\cap U_j}(s_j)$, there is a unique $s {\in} F(U)$ with
$\rho^{U}_{U_i}(s){=}s_i$ for all $i$.
\end{enumerate}
\end{quote}
The morphisms of (pre)sheaves are natural transformations.
We denote by ${\sf PreSh}(I)$ the category of presheaves over $I$ and 
by ${\sf Sh}(I)$ the category of sheaves over $I$. 
\label{top-sh} 
\end{definition}

\begin{definition}
The {\em stalk} of a sheaf $F$ on $I$ at a point $i \in I$
is the colimit $F_i = \limright_{i \in U} F(U)$, where $U$ ranges over all
open neighborhoods of $i$. 
The assignment $F \mapsto F_i$ defines the stalk functor at $i$, 
${\sf Stalk}_i : {\sf Sh}(I) \rightarrow {\sf Set}$. 
\end{definition}

\noindent Sheaves can be defined also in a different way. 
An indexed system\index{indexed system} 
of sets $(F_i)_{i \in I}$ can alternatively be
regarded as a map $f : F =\coprod_{i \in I} F_i \rightarrow I$, with
the property that for every $x \in F$, $f(x) = i$ if and only if $x
\in F_i$. If the index set $I$ has a topology, then the set $F$ can be endowed
with a topology such that $f$ is continuous (i.e.\ 
the sets in the family $(F_i)_{i \in I}$ are 
continuously indexed).

\begin{definition} A {\em bundle}\index{bundle}
over $I$ is a
triple $(F, f, I)$ where $F$ and $I$ are topological  spaces and 
$f : F \rightarrow I$ is continuous.  
For every $i \in I$, $f^{-1}(i)$ will be denoted by $F_i$. Then 
$F = \coprod_{i \in I} F_i$.
Let $(F, f, I)$ and $(G, g, I)$ be two bundles
over $I$. A morphism\index{bundles, morphism} 
between $(F, f, I)$ and $(G, g, I)$ is a continuous map $h :
F \rightarrow G$ such that $g \circ h = f$.

\noindent 
The {\em category of bundles} over $I$ is denoted {\sf Sp/}$I$.
\end{definition} 

\noindent 
Let {\sf LH}$/I$ be the full subcategory of {\sf Sp}$/I$ with objects
$(F, f, I)$,  where $f : F \rightarrow I$ a local
homeomorphism (i.e.\ for every $a \in F$ there are open neighborhoods
$U$ and $U'$ of $a$ respectively $f(a)$ such that $f : U \rightarrow
U'$ is a homeomorphism).

\begin{definition}
Let $(F, f, I)$ be a bundle over
$I$.  A {\em partial section}\index{section, partial} defined on a 
open subset $U \subseteq I$
is a continuous map $s : U \rightarrow F$ with the property that $f
\circ s$ is the inclusion $U \subseteq I$.  A section defined on $I$
is called {\em global section}\index{section, global}.  
The set of all partial sections over
the open subset $U$ of $I$ will be denoted by $\Gamma(F, f)(U)$.
\end{definition}
The following links between (pre)sheaves and bundles exist:
\begin{itemize}
\item For every bundle $(F, f, I)$ let 
$\Gamma(F) = \{ s : I \rightarrow F \mid s \mbox{ continuous and } 
f \circ s = id_I \}$, the set of all global sections of $F$.
This defines a functor $$\Gamma : {\sf Sp}/I \rightarrow {\sf PreSh}(I).$$

\item Let $F$ be a presheaf on $I$. 
For every $i \in I$ let $F_i$ be the 
stalk of $F$ at a point $i \in I$. 
The collection of stalks $(F_i)_{i \in I}$ is an $I$-indexed family of
sets.  Let $D(F)$ denote the disjoint union of the stalks, and let
$\pi : D(F) \rightarrow I$ be the canonical projection on $I$ defined by 
$\pi(x) = i$ iff $x \in F_i$.  
For $s \in F(U)$ and $i \in U$, let $s_i$ be the image of $s$ in $F_i$. 
The map $\overline{s} : U \rightarrow D(F)$,
$\overline{s}(i) = s_i$ defines a partial section of 
$\pi : D(F) \rightarrow I$; we impose on $D(F)$ the coarsest topology for
which all such sections are continuous.
$D(F) = (D(F), \pi, I)$ is a bundle. 
This construction defines a functor 
$$D : {\sf PreSh}(I) \rightarrow {\sf Sp}/I.$$ 
\end{itemize}
\begin{theorem}[cf.\ \cite{Johnstone82,MacLaneMoerdijk92}] 
The functor 
$D :$ {\sf PreSh}$(I) \rightarrow $ {\sf Sp}$/I$ 
preserves finite limits and is left adjoint to $\Gamma :$ {\sf
Sp}$/I \rightarrow $ {\sf PreSh}$(I)$.  
The functors $D, \Gamma$ restrict to an
equivalence of categories between {\sf Sh}$(I)$ and {\sf LH}$/I$.
\end{theorem}
$\Gamma \circ D : {\sf PreSh}(X) \rightarrow 
{\sf Sh}(X)$ is known as the {\em sheafification 
functor}. 
\begin{theorem}[cf.\ \cite{Johnstone82,MacLaneMoerdijk92}] 
The inclusion ${\sf Sh}(X) {\rightarrow} {\sf PreSh}(X)$ has a left adjoint,
$\Gamma {\circ} D : {\sf PreSh}(X) {\rightarrow} 
{\sf Sh}(X)$. 
The sheafification functor $\Gamma {\circ} D$ 
preserves 
all finite limits.
\end{theorem}

\section{States, partial actions}

Let ${\sf InSys}$ be a family of  systems satisfying 
conditions (i), (ii), (iii) in Section~\ref{insys}, and 
$\Omega({\sf InSys})$ be the topology on ${\sf InSys}$ 
consisting of all subsets 
{\sf InSys} which are closed under t.c.\ subsystems.
We define functors modeling states and parallel actions: 
\begin{description}
\item[${\sf (St)}$] 
${\sf St} : \Omega({\sf InSys})^{\mbox{\scriptsize op}} \rightarrow {\sf Set}$
is defined as follows:   

\medskip
{\em Objects:} ${\sf St}(U) = \{ (s_i)_{S_i \in U} \mid s_i \in St(S_i),
\mbox{ and if } S_i \hookrightarrow S_j \mbox{ then } s_i = {s_j}_{|X_i} \}$;\\[-2.2ex]

{\em Morphisms:} if $U_1 {\stackrel{\iota}{\subseteq}} U_2$,
${\sf St}(\iota) {:} {\sf St}(U_2) {\rightarrow}
{\sf St}(U_1)$ is 
${\sf St}(\iota)((s_i)_{S_i \in U_2}) {=} (s_i)_{S_i \in U_1}$.

\bigskip

\item[${\sf (Pa)}$] 
${\sf Pa} : \Omega({\sf InSys})^{\mbox{\scriptsize op}} \rightarrow {\sf Set}$
is defined as follows:  

\medskip
{\em Objects:}
${\sf Pa}(U) = \{ (f_i)_{S_i \in U} \mid f_i \in Pa(S_i),
\mbox{ and if } S_i \hookrightarrow S_j \mbox{ then } f_i = {f_j}_{|A_i} \}$;\\[-2.2ex]

{\em Morphisms:} if $U_1 {\stackrel{\iota}{\subseteq}} U_2$,
${\sf Pa}(\iota) {:}
{\sf Pa}(U_2) {\rightarrow} {\sf Pa}(U_1)$ is
${\sf Pa}(\iota)((f_i)_{S_i \in U_2}) {=} (f_i)_{S_i \in U_1}$.
\end{description}

\begin{example}
\label{ex-states-sheaf}
Consider the 
family ${\sf InSys} = \{ S_1, S_{12}, S_2 \}$ in 
Example~\ref{ex-colimit}. 

\smallskip
\noindent {\em States.} Any tuple $(s_1, s_2, s_{12})$, 
where $s_i \in St(S_i)$ for $i \in \{ 1, 2, 12 \}$
and ${s_1}_{|X_{12}} = {s_2}_{|X_{12}} = 
s_{12}$, is an element in ${\sf St}({\sf InSys})$. 
Assume first that $k \leq l$.  
\begin{itemize}
\item Let $s_i : X_{S_i} \rightarrow M$ 
be such that $s({\sf TrainIndex}_i) = i$ for all $i \in \{ 1, \dots, l \}$, 
and such that ${s_1}_{|X_{12}} = {s_2}_{|X_{12}} = 
s_{12}$. Then $(s_1, s_2, s_{12}) \in {\sf St}({\sf InSys})$. 
\item Let $s_1 : X_{S_1} \rightarrow M$ be defined by 
$s({\sf TrainIndex}_i) = i$ for all $i \in \{ 1, \dots, l \}$, and 
$s_2 : X_{S_2} \rightarrow M$ be defined by 
$s({\sf TrainIndex}_i) = i+1$ for all $i \in \{ k, \dots, n \}$. 
$s_1 \in St(S_1)$, $s_2 \in St(S_2)$, but they do not agree on 
the common control variables (in particular, $s_1({\sf TrainIndex}_k) = k, 
s_2({\sf TrainIndex}_k) = k+1$). 
So $(s_1, s_2, {s_1}_{|X_{S_{12}}}) \not\in {\sf St}({\sf InSys})$. 
\end{itemize}
Assume now that $l < k$. Then $S_{12}$ is the system with an empty 
set of control variables. Hence, $s_1 : X_{S_1} \rightarrow M$ defined by 
$s({\sf TrainIndex}_i) = i$ for all $i \in \{ 1, \dots, l \}$, and 
$s_2 : X_{S_2} \rightarrow M$, defined by 
$s({\sf TrainIndex}_i) = i+1$ for all $i \in \{ k, \dots, n \}$, 
agree on the common variables. Therefore 
$(s_1, s_2, {s_1}_{|X_{S_{12}}}) \in {\sf St}({\sf InSys})$.

\smallskip
\noindent 
Let $U = \{ S_1, S_{12}, S_2 \}$ and $U_1 = \{ S_1, S_{12} \}$ be the 
two sets in $\Omega({\sf InSys})$ which contain $S_1$, and
let $i$ be the inclusion between $U_1$ and $U$. 
Then ${\sf St}(i) : {\sf St}(U) \rightarrow {\sf St}(U_1)$ is  
defined by $St(i)(s_1, s_2, s_{12}) = \rho^{U}_{U_1}(s_1, s_2, s_{12}) = 
(s_1, s_{12})$. 

\medskip
\noindent {\em Parallel Actions.} Any tuple $(f_1, f_2, f_{12})$, 
where $f_i \in Pa(S_i)$ for $i \in \{ 1, 2, 12 \}$
and ${f_1}_{|A_{12}} = {f_2}_{|A_{12}} = 
f_{12}$, is an element in ${\sf Pa}({\sf InSys})$. 
In particular: 
\begin{itemize}
\item $(f_1, f_2, f_{12})$ with 
$f_j^{-1}(1) = \{ {\sf report_i} \mid i \in I_j \} \cup {\sf update}$.
These are admissible parallel actions in the corresponding systems, 
and ${f_1}_{|A_{12}} = {f_2}_{|A_{12}} = 
f_{12}$. Then $(f_1, f_2, f_{12}) \in {\sf Pa}({\sf InSys})$. 
\end{itemize}
Tuples $(f_1, f_2, f_{12})$ which do not satisfy these 
conditions are not in ${\sf Pa}({\sf InSys})$:
\begin{itemize}
\item $(f_1, f_2, f_{12})$ with 
$f_j^{-1}(1) = \{ {\sf report_i} \mid i \in I_j\} \cup {\sf update} \cup 
\{ {\sf move}_i \mid i \in I_j \}$ is not in 
${\sf Pa}({\sf InSys})$, because the components are not admissible parallel actions.

\item $(f_1, f_2, f_{12})$ with $f_1^{-1}(1) = 
\{ {\sf report_i} \mid i \in I_1 \} \cup {\sf update}$ and 
$f_2^{-1}(1) =  \{ {\sf move}_i \mid i \in I_2 \}$ is not in 
${\sf Pa}({\sf InSys})$, because the components do not agree on 
$A_{12}$.
\end{itemize}
\end{example}

\begin{theorem}[\cite{Sofronie-Stokkermans99}]
The functors ${\sf St}$ and ${\sf Pa}$ are sheaves on 
${\sf InSys}$. 
For each $S_i {\in} {\sf InSys}$, 
the stalk at $S_i$ of ${\sf St}$ (resp.\ ${\sf Pa}$)
is in bijection with $St(S_i)$  (resp.\ $Pa(S_i)$). 
Moreover, for each $U \in \Omega({\sf InSys})$, 
${\sf St}(U)$ (resp.\ ${\sf Pa}(U)$) is 
in bijection with $St(S_U)$ (resp.\ $Pa(S_U)$), 
where $S_U$ is the colimit of the diagram defined by $U$.
\end{theorem}

\begin{example}
\label{ex-colimits-sheaf}
Let 
${\sf InSys} = \{ S_1, S_{12}, S_2 \}$ as defined in 
Example~\ref{interconnect-3sys} (with $k \leq l$): 
\begin{enumerate}
\item[(1)] An example of an open cover for 
$U = \{ S_1, S_2, S_{12} \}$ is $\{ U_1, U_2, U_{12} \}$, where   
$U_1 = \{ S_1, S_{12} \}, U_2 = \{ S_2, S_{12} \}, U_{12} = \{ S_{12} \}$.
Let $(s_1, s_{12}) \in St(U_1)$ and $(t_2, t_{12}) \in St(U_2)$ be 
such that $\rho^{U_1}_{U_{12}}(s_1, s_{12}) = 
\rho^{U_2}_{U_{12}}(t_2, t_{12})$. Then $s_{12} = t_{12}$ and there is a 
unique element $(s_1, t_2, s_{12}) \in St(U)$ such that 
$\rho^{U}_{U_1}(s_1, t_2, s_{12}) = (s_1, s_{12})$ and 
$\rho^{U}_{U_2}(s_1, t_2, s_{12}) = (t_2, t_{12})$. 
 Similar for ${\sf Pa}$.
\item[(2)] The stalk of ${\sf St}$ at $S_1$ is the colimit 
of the diagram 
${\sf St}(U) {\stackrel{{{\sf St}(i)}}{\rightarrow}}
{\sf St}(U_1) {\stackrel{{{\sf St}(id)}}{\rightarrow}}
{\sf St}(U_1)$ and hence in bijection with 
${\sf St}(U_1)$.  Similarly for ${\sf Pa}$.  
\item[(3)] 
It can be seen that 
${\sf St}(U)$ is in bijection with $St(S)$, where $S$ is the system in 
the example in Section~\ref{example-system}:  
Let $(s_1, s_2, s_{12}) \in {\sf St}(U)$. Then $s : X \rightarrow M$ 
defined by $s(x) = s_i(x)$ iff $x \in X_i$ is well defined (due to the 
definition of ${\sf St}(U)$) and in $St(S)$. 
Conversely, if $s \in St(S)$, then $(s_{X_1}, s_{X_2}, s_{|X_{12}}) \in 
{\sf St}(U)$.

Also ${\sf Pa}(U)$ is in bijection with $Pa(S)$: 
If $(f_1, f_2, f_{12}) \in {\sf Pa}(U)$ then $f : A \rightarrow \{ 0, 1 \}$ 
defined by $f(x) = f_i(x)$ iff $x \in A_i$ is well defined (due to the 
definition of ${\sf Pa}(U)$). It can also be checked 
that if $f_1 \models C_1$ and $f_2 \models C_2$ then $f \models C$. 
Thus, $f \in Pa(S)$. The converse is immediate.
\end{enumerate}
Assume now that ${S_1, S_2, S_{12}}$ are as in Example~\ref{ex-colimit}
but $l < k$, say $l = k-1$. The open cover and stalk 
construction in (1) and (2) above are the same. However, 
${\sf St}(U)$ is in bijection with $St({\overline S})$, where 
${\overline S}$ is the colimit of the diagram defined by $U$ 
as described in Example~\ref{ex-colimit} which in this case is 
different from $S$. In particular, $s: X \rightarrow M$ with 
$s({\sf TrainIndex}_1) = 1, s({\sf TrainIndex}_2) = 2, \dots, 
s({\sf TrainIndex}_{k-1}) = k-1$ and  
$s({\sf TrainIndex}_k) = k+1, \dots,  s({\sf TrainIndex}_{n-1}) = n$ 
is a state of ${\overline S}$, but not of $S$. 
\end{example}

\section{Transitions}

Let ${\sf InSys}$ be a family of  systems satisfying 
conditions (i), (ii), (iii) in Section~\ref{insys}.
We define a functor modeling transitions: 
\begin{description}
\item[${\sf (Tr)}$] 
${\sf Tr}:\Omega({\sf InSys})^{\mbox{\scriptsize op}} \rightarrow {\sf Set}$ 
is defined as follows:  

\medskip
{\em Objects:} 
${\sf Tr}(U) = \{(f,s,s') \mid f=(f_i)_{S_i \in U} \in {\sf Pa}(U),
s = (s_i)_{S_i \in U} \in {\sf St}(U),$ \\
$\mbox{\hspace{3cm}} s'=(s'_i)_{S_i \in U} \in {\sf St}(U),
(s_i, s'_i) \in Tr_{S_i}(f_i),$ for all $S_i \in U\};$ 

\begin{tabular}{@{}ll}
{\em Morphisms:} if $U_1 \stackrel{\iota}{\subseteq} U_2$, & 
${\sf Tr}(\iota) : {\sf Tr}(U_2) \rightarrow
{\sf Tr}(U_1)$ is defined by \\
& ${\sf Tr}(\iota)((f, s, s')) = 
({\sf Pa}(\iota)(f), {\sf St}(\iota)(s), {\sf St}(\iota)(s'))$, 
\end{tabular}

\medskip
where, for every $S_i$ in {\sf InSys} and
$f_i \in Pa(S_i)$, $Tr_{S_i}(f_i)$ is the
transition relation associated
to $f_i$ in $S_i$ as explained in Section~3.
\end{description}

\begin{example}
Consider the family $\{ S_1, S_{12}, S_2 \}$ in 
Example~\ref{interconnect-3sys}.
With the notation introduced in Example~\ref{interconnect-3sys}, let: 
\begin{itemize}
\item $s_j({\sf ActualPos}_i) = a_i,$ 
$s_j({\sf RepPos}_i) = r_i, s_j({\sf Mode}_i) = m_i$, 
for $i \in I_j$; 
\item $f_j$ be such that 
$f_j^{-1}(1) = \{ {\sf report_i} \mid i \in I_j \} \cup {\sf update}$, and 
\item $s'_j$ be defined by: 
$s'_j({\sf ActualPos}_i) \,{=}\, a_i, s'_j({\sf RepPos}_i) \,{=}\, a_i, 
s'_j({\sf Mode}_i) \,{=}\, m'_i$, where $m'_i$ is computed according to the 
transition rules for ${\sf update}$ in Example~\ref{example-system}. 
\end{itemize}
\begin{tabular}{@{}ll}
Then: &  $f_i \in Pa(S_i), s_i, s'_i \in St(S_i), (s_i, s'_i) \in Tr(S_i)$ 
for $i \in \{ 1, 2, 12 \}$,\\
&  ${f_1}_{|A_{12}} = {f_2}_{|A_{12}} = f_{12}$ and  
${s_1}_{|X_{12}} = {s_2}_{|X_{12}} = {s_{12}}$.
\end{tabular}
 
\medskip
\noindent Hence,  
$((f_1, s_1, s'_1), (f_2, s_2, s'_2), (f_{12}, s_{12}, s'_{12}))$ 
is in ${\sf Tr}({\sf InSys})$.
\end{example}

\begin{theorem}[\cite{Sofronie-Stokkermans99}]
The functor 
${\sf Tr}:\Omega({\sf InSys})^{\mbox{\scriptsize op}} \rightarrow {\sf Set}$
is a subsheaf of ${\sf Pa} \times {\sf St} \times {\sf St}$.
Moreover: 
\begin{itemize}
\item For every $S_i \in {\sf InSys}$, 
the stalk of ${\sf Tr}$ at $S_i$
is in bijection with $Tr(S_i) = \{ (f,s,s') \mid 
(s, s') \in Tr_{S_i}(f)\}$. 
\item If the transitions obey either
({\bf Disj}) or ({\bf  Indep}), then, 
for every $U \in \Omega({\sf InSys})$, 
${\sf Tr}(U)$ is in bijection with  $Tr(S_U) = \{ (f, s, s')
\mid (s, s') \in Tr_{S_U}(f) \}$, 
where $S_U$ is the colimit of the diagram defined by $U$.
\end{itemize}\label{thm1}
\end{theorem}

\begin{example}
Consider the family 
$\{ S_1, S_{12}, S_2 \}$ in Example~\ref{interconnect-3sys}.
Consider the transition $((f_1,s_1,s'_1), (f_2,s_2,s'_2), (f_{12},s_{12},s'_{12})) 
\in {\sf Tr}(U)$. Let $f : A \rightarrow \{ 0, 1 \}$ 
be defined by $f(x) = f_i(x)$ iff $x \in A_i$ is well defined. 
Then $f \in Pa(S)$. 
Similarly, $s, s' : X \rightarrow M$, defined by 
($s(x) = s_i(x)$ and $s'(x) = s'_i(x)$) iff $x \in X_i$ are 
well defined and in $St(S)$.

As shown in Example~\ref{example-disj}, 
the transitions in all systems $S_1, S_2, S_{12}$ obey 
condition $({\bf Disj})$. The changes of the 
components of parallel actions are not contradictory and 
affect only the variables the actions depend upon. Thus, $(s, s')$ 
is in the transition induced (according to rule  $({\bf Disj})$) by $f$.
Hence, $(s, s') \in Tr_S(f)$. 
The converse is an immediate consequence of the fact that, as showed in 
Example~\ref{example-transition-connected}, 
$S_1, S_2, S_{12}$ are transition-connected subsystems of $S$.
\end{example}

\section{Behavior in time}
In \cite{Goguen92}, the behavior of a given 
system $S$ in time is modeled  by a functor 
$F : {\cal T}^{op} \rightarrow {\sf Set}$, 
where  ${\cal T}$ is the basis for the topology on 
${\mathbb N}$ 
consisting of all the sets $\{ 0, 1, \dots, n \}, n \in {\mathbb N}$.
Intuitively, for every $T \in {\cal T}$,  
$F(T)$ represents the succession of the states of the systems ``observed'' 
during the interval of time $T$.
We analyze various alternative possibilities of modeling behavior.
\subsection{Behavior as successions of states and actions}
Since we are interested in actions as well as states, 
we present a different description of behavior. 
Let  ${\cal T}$ consist of ${\mathbb N}$ together with 
all sets $\{ 0, 1, \dots, n \}, n \in {\mathbb N}$.
The behavior in an interval $T \in {\cal T}$ of a complex system  
obtained by interconnecting a family ${\sf InSys}$ (satisfying 
conditions (i)--(iii) in Section~\ref{insys})
is modeled by all successions of pairs (state, action) of the component 
subsystems that can be observed during $T$, i.e.\ by the functor
${\sf B}_T : \Omega({\sf InSys})^{\mbox{\scriptsize op}} 
\rightarrow  {\sf Set}$ defined as follows:

\medskip
{\em Objects:}~~~~~ for $U \in \Omega({\sf InSys})$, 
${\sf B}_T(U) {=} \{ h {:} T {\rightarrow} {\sf St}(U) {\times} {\sf Pa}(U)  
{\mid} K(h, T) \}$,

{\em Morphisms:} for $U_1 \stackrel{\iota}{\subseteq} U_2$ by 
${\sf B}_T(\iota) {:} 
{\sf B}_T(U_2) {\rightarrow} {\sf B}_T(U_1)$,  
where if  
$h \in {\sf B}_T(U_2)$,

$~~~{\sf B}_T(\iota)(h) {=} ({\sf St}(\iota) {\times} 
{\sf Pa}(\iota)) {\circ} h 
: T \stackrel{h}{\longrightarrow} {\sf St}(U_2) {\times} {\sf Pa}(U_2)
\stackrel{{\small {\sf St}(\iota) {\times} {\sf Pa}(\iota)}}{-\!\!\!-\!\!\!-\!\!\!-\!\!\!-\!\!\!\longrightarrow}
{\sf St}(U_1) {\times} {\sf Pa}(U_1)$.

\medskip
\noindent 
Here $K(h, T)$ expresses the fact that for every $n$, if
$n, n+1 \in T$ and $h(n) = (s, f)$, $h(n+1) = (s', f')$ then
$(f,s,s') \in {\sf Tr}(U)$. 

\begin{example}
We illustrate the definition above. 
Let $T = {\mathbb N}$, and let $U = \{ S_1, S_2, S_{12} \}$ 
as in Example~\ref{interconnect-3sys}.
We represent an element $h$ in ${\sf B}_T({\sf InSys})$ as a table
(first row: arguments $i$ of $h$, second row: the value $h(i)$, i.e.\ a pair of tuples): 

\medskip
{\small 
$$\begin{array}{@{}|@{}c@{}||@{}c@{}|@{}c@{}|@{}c@{}|@{}c@{}|@{}c@{}|@{}c@{}|@{}c@{}|@{}c@{}|@{}c@{}|@{}c@{}|@{}c@{}|@{}c@{}|@{}c@{}|@{}c@{}|@{}c@{}|@{}c@{}|@{}c@{}|@{}c@{}|}
\hline 
 & \multicolumn{18}{c|}{h(i)} \\
\cline{2-19} 
  & \multicolumn{9}{c|}{{\sf St}(U)} & \multicolumn{9}{c|}{{\sf Pa}(U)} \\
\cline{2-19}
i  & \multicolumn{3}{c|}{{\sf St}(S_1)} & \multicolumn{3}{c|}{{\sf St}(S_{12})} & \multicolumn{3}{c|}{{\sf St}(S_2)} & \multicolumn{3}{c|}{{\sf Pa}(S_1)} & \multicolumn{3}{c|}{{\sf Pa}(S_{12})} & \multicolumn{3}{c|}{{\sf Pa}(S_2)} \\
\cline{2-19}
& \multicolumn{3}{c|}{(i \in I_1)} & \multicolumn{3}{c|}{(i \in I_{12})} & \multicolumn{3}{c|}{(i \in I_2)} & \multicolumn{3}{c|}{(i \in I_1)} & \multicolumn{3}{c|}{(i \in I_{12})} & \multicolumn{3}{c|}{(i \in I_2)} \\
\cline{2-19}
& {\sf ActPos}_i & {\sf RepPos}_i & {\sf Mode}_i & \multicolumn{3}{c|}{\text{ (restr.) }} & {\sf ActPos}_i & {\sf RepPos}_i & {\sf Mode}_i & {\sf rep}_i & {\sf upd} & {\sf move}_i &  \multicolumn{3}{c|}{\text{ (restr.) }} & {\sf rep}_i & {\sf upd} & {\sf move}_i \\
\hline 
0 & a_i & r_i & m_i & a_i & r_i & m_i & a_i & r_i & m_i & 1 & 1 & 0 & 1 & 1 & 0 & 1 & 1 & 0 \\
1 & a_i & a_i & m'_i & a_i & a_i & m'_i & a_i & a_i & m'_i & 0 & 0 & 1 & 0 & 0 & 1 & 0 & 0 & 1 \\
2 & a'_i & a_i & m'_i & a'_i & a_i & m'_i & a'_i & a_i & m'_i & 1 & 1 & 0 & 1 & 1 & 0 & 1 & 1 & 0 \\
3 & a'_i & a'_i & m''_i & a'_i & a'_i & m''_i & a'_i & a'_i & m''_i & 1 & 1 & 0 & 1 & 1 & 0 & 1 & 1 & 0 \\

\dots & ... & ... & ... & ... & ... & ... & ... & ... & ... & ... & ... & ... & ... & ... & ... & ... & ... & ...  \\
& & & & & & & & & & & & & & & & & &   \\
\hline 
\end{array}$$
} 
\end{example}

\medskip
\begin{theorem}[\cite{Sofronie-Stokkermans99}]
 Let $B_T(S) = \{ h : T \rightarrow St(S) \times Pa(S)  \mid  K_S(h, T) \}$,
where $K_S(h, T)$ expresses the fact that for every $n$, if
$n, n+1 \in T$ and $h(n) = (s, f)$, $h(n+1) = (s', f')$ then
$(s,s') \in Tr_{S}(f)$. Then: 
\begin{itemize} 
\item For every  $T \in {\cal T}$, 
${\sf B}_T : \Omega({\sf InSys})^{\mbox{\scriptsize op}} \rightarrow {\sf Set}$
is a sheaf.
\item For every $S_i \in {\sf InSys}$, 
the stalk at $S_i$ is in bijection with $B_T(S_i)$. 
\item If the transitions obey 
({\bf Disj}) or ({\bf Indep}), then, 
for every $U \in \Omega({\sf InSys})$, 
${\sf B}_T(U)$ is in bijection with $B_T(S_U)$, where 
$S_U$ is the colimit of the diagram defined by $U$. 
\end{itemize}\label{thm3}
\end{theorem}

\subsection{Behavior: Admissible Parallel Actions as Words}  
If we ignore the states, the behavior of any system $S$ can be expressed by 
a subset $L_S$ of the free monoid $Pa(S)^*$ 
over the set of possible actions of $S$, where: 
\begin{eqnarray*}
L_S & =  \{ f_1 \dots f_n \mid & \exists h : \{ 0, \dots, n \} \rightarrow St(S) \times Pa(S),  \exists s_i \in St(S), \mbox{ s.t. } \\
& & \forall i \in \{ 0, \dots, n-1 \}, (s_i, s_{i+1}) \in Tr_S(f_i) \} \subseteq Pa(S)^*.
\end{eqnarray*}
Consider the family $\{ Pa(S_i)^* \mid S_i \in {\sf InSys} \}$.
If $S_i, S_j \in {\sf InSys}$ and $S_i \hookrightarrow S_j$, 
let $\rho^{S_j}_{S_i} : Pa(S_j) \rightarrow Pa(S_i)$ be the restriction
to $S_i$. The restriction extends 
to a homomorphism of monoids, 
$p^j_i : Pa(S_j)^* {\rightarrow} Pa(S_i)^*$.
If there is no risk of 
confusion, in what follows we will abbreviate 
$p^j_i(w_j)$ by ${w_j}_{|S_i}$. 
Let $M({\sf InSys})$ be defined by:

\vspace{-6mm}
$$M({\sf InSys}) = \{ (w_i)_{S_i \in {\sf InSys}} \mid w_i \in Pa(S_i)^* 
\mbox{ and } \forall S_i \hookrightarrow S_j, p^j_i(w_j) = w_i \}.$$ 

\vspace{-2mm}
\noindent  It can be seen that 
$M({\sf InSys})$ 
is the limit of the diagram $\{ Pa(S_i)^* \mid S_i \in {\sf InSys} \}$
(with the morphisms $p^j_i$ for every $S_i \hookrightarrow S_j$).

\begin{theorem}
Let $M : \Omega({\sf InSys})^{op} \rightarrow {\sf Sets}$ 
be defined as follows:

{\em Objects:}~~~~ $M(U) = \{ (w_i)_{S_i {\in} V} \mid  w_i {\in} Pa(S_i)^*,  
{w_i}_{|S_j} = w_j \mbox{ for every } S_j \hookrightarrow S_i \},$

{\em Morphisms:} 
if $\iota : U_1  \subseteq U_2$,  
$M(\iota)  : M(U_2) \rightarrow M(U_1)$ is defined for every 

$~~~~~~~~~~~~~~~~~~~~~~~~~~~~~~~~~~~~~(w_i)_{S_i \in U_2}$ by 
$ M(\iota)((w_i)_{S_i \in U_2}) = (w_i)_{S_i \in U_1}$.

\noindent Then $M$ is a sheaf of monoids.
$M(V)$ is the limit of the diagram 
$\{ Pa(S_i)^* \mid S_i \in V \}$ (with morphisms $p^j_i : Pa(S_j)^* {\rightarrow} Pa(S_i)^*$ whenever $S_i \hookrightarrow S_j$). 
\label{proof-sheaf-of-monoids}
\end{theorem}
\Proof Let $U \in \Omega({\sf InSys})$ and 
$\{ U_k \mid k \in K \}$ be a cover for $U$. 
Let $\{ w_k \}_{k \in K}$ be a family of elements,
such that for every $k \in K$, 
$w_k = (w^i_k)_{S_i \in U_k}$ and 
for every $k_1, k_2 \in K$, if $S_i \in U_{k_1} \cap U_{k_2}$ 
then $w^i_{k_1} = w^i_{k_2}$. 

We define $w = (w_i)_{Si \in U}$ as follows: 
for every $S_i \in U$, $S_i \in U_k$ for some $k$. Then $w_i$ is defined 
to be  $w^i_k$. Note that $w_i$ is well defined because of the 
compatibility of the family 
 $\{ w_k \}_{k \in K}$, and $p^U_{U_k}(w) = w_k$ for every 
$k \in K$. The uniqueness of $w$ follows from the fact that 
for every $w' = (w'_i)_{S_i \in U}$ such that 
$p^U_{U_k}(w') = w_k$ for every 
$k \in K$ we have $w'_i = w^k_i$ for every $S_i \in U_k$. 

The fact that 
$M(V)$ is the limit of the diagram 
$\{ Pa(S_i)^* \mid S_i \in V \}$ (with the corresponding morphisms) can be 
checked without difficulty. \QED

\medskip
\noindent 
{\bf Remark:} Let $S$ be 
the colimit of the diagram defined by 
$U$. The connection between $Pa(S)^*$ and $M(U)$ is rather loose: 
Let  $p : Pa(S)^* \rightarrow M(U)$ be defined by 
$p(f_1 \dots f_n) = ((f_1 \dots f_n)_{|S_i})_{S_i \in U} \in M(U)$.
If we identify the empty action
with the empty word $\epsilon$, $p$ may not be injective 
as can be seen from the following example: 

\begin{example}
\label{ex-inj}
Let $S_1$ and $S_2$ be as defined in Example~\ref{interconnect-3sys}, 
where trains are 
indexed by $I_1 = \{ k_1, \dots, n \}$ and $I_2 = \{ 1, \dots, k_2 \}$ 
and $k_2 < k_1$, with the difference that ${\sf update}$ is omitted as 
in Example~\ref{example-indep}. 
Let ${\sf InSys} = \{ S_1, S_2, \emptyset \}$. 
Let $w_1 = f_1 f_2$ and $w_2 = f_2 f_1$, where $f_1^{-1}(1) = \{ {\sf report}_i \mid i \in I_1 \}$  and 
$f_2^{-1}(1) = \{ {\sf move}_j \mid j \in I_2 \}$. 
Note that ${f_1}^{-1}_{|A_1}(1) = \{ {\sf report}_i \mid i \in I_1 \}$, 
${f_2}^{-1}_{|A_1}(1) {=} {f_1}^{-1}_{|A_2}(1) {=} \emptyset$, and 
${f_2}^{-1}_{|A_2}(1) = \{ {\sf move}_j \mid j \in I_2 \}$. 
Thus, 
\begin{eqnarray*}
p(w_1) & = & ((f_1 f_2)_{|S_1}, (f_1 f_2)_{|S_2}, (f_1 f_2)_{|\emptyset}) 
       ~ = ~ 
(({f_1}_{|A_1} {f_2}_{|A_1}), ({f_1}_{|A_2} {f_2}_{|A_2}), \epsilon) \\
& = & 
        (f_1 \epsilon, \epsilon f_2, \epsilon) = (\epsilon f_1, f_2 \epsilon, \epsilon) 
~ =~ (({f_2}_{|A_1} {f_1}_{|A_1}), ({f_2}_{|A_2} {f_1}_{|A_2}), \epsilon) 
~ = ~ p(w_2),
\end{eqnarray*} 
but $w_1 \neq w_2$. 
\end{example}

\noindent The next example shows that 
$p : Pa(S)^* \rightarrow M(U)$ is not necessarily 
onto: 
There may exist compatible families 
(even if we only consider singleton parallel actions) 
of sequences of actions that cannot be ``glued together'' to a sequence 
of actions on $Pa(S)$.  A similar result appears in 
\cite{MonteiroPereira86} (in that 
case, no parallelism is allowed). 

\begin{example}
Let $S_1, S_2, S_3$ be three systems all having  the same language, 
the same constraints on variables and the same model for the variables, 
such that 

$\begin{array}{lll}
A_{S_1} = \{ a, b, d \}, & A_{S_2} = \{ b, c, e \}, & A_{S_3} = \{ a, c, f \} \\
C_{S_1} = \{ a \wedge b = 0 \}~~ 
& C_{S_2} = \{ b \wedge c = 0 \}~~ 
& C_{S_3} = \{ a \wedge c = 0 \} 
\end{array}$

\noindent Let $S$ be the system obtained by interconnecting the systems 
$S_1, S_2, S_3$. 
Then 
$A_{S} = \{ a, b, c, d, e, f \}, C_{S} = \{ a \wedge b = 0 , b \wedge c = 0, 
a \wedge c = 0 \}.$
Consider $w_1 = a b \in Pa(S_1)^*$, $w_2 = b c \in Pa(S_2)^*$,
$w_3 = c a \in Pa(S_3)^*$. 
It is easy to see that $p^1_{12}(w_1) = p^2_{12}(w_2) = b$, 
$p^2_{23}(w_2) = p^3_{23}(w_3) = c$, $p^1_{13}(w_1) = p^3_{13}(w_3) = a$, 
but there is no $w \in Pa(S)^*$ such that $w_{|S_i} = w_i, i = 1, 2, 3$.
\label{ex-onto}
\end{example}

\noindent We investigate therefore other ways of modelling behavior for which 
tighter links between local and global behavior exist.

\subsection{Behavior: Partially Commutative Monoids}
In what follows we assume that the constraints on actions
are all of the form $a_i \wedge a_j = 0$ (they state which actions 
cannot be performed in parallel).

\begin{definition}
Let $S$ be a system with the property that the constraints on actions
are all of the form $a_i \wedge a_j = 0$. 
The {\em dependence graph} of $S$ is the graph $(A_S, D_S)$ having 
as set of vertices $A_S$, and where $D_S$ is defined by  
$(a_1, a_2) \in D_S$ if 
$a_1 = a_2$ or $a_1 \wedge a_2 = 0 \in C_S$. 
\end{definition}

\noindent 
For every system $S$ with dependence graph $(A_S, D_S)$ we denote by 
$M(S) = M(A_S, D_S)$ the free partially commutative monoid defined by 
$(A_S, D_S)$, i.e.\ the quotient of $A_S^*$ by the congruence relation 
generated  by $a_1 a_2 = a_2 a_1$ for every 
$(a_1, a_2) \in (A_S \times A_S) \backslash D_S$.
For basic properties of (free) partially commutative monoids
we refer e.g.\ to \cite{Diekert90}, pp.9-29 and 67-79.

\medskip
\noindent For every $S_i \in {\sf InSys} \backslash \emptyset$, 
let $M(S_i) = A_{S_i}^*/\theta_i$ (where $\theta_i$ is the congruence 
defined as explained above from $(A_{S_i} \times A_{S_i}) \backslash D_{S_i}$) 
be the partially 
commutative monoid associated with the dependence graph of $S_i$. 
Let $S$ be the colimit of the diagram defined by ${\sf InSys}$. 
Then $A_S = \bigcup_{S_i \in {\sf InSys}} A_i$ and 
$D_S =  \bigcup_{S_i \in {\sf InSys}} D_i$. 
Hence, for every $S_i \in {\sf InSys}$ 
there is a canonical projection 
$p_i : M(S) \rightarrow M(S_i)$ which is onto. Let $ker(p_i)$ 
be the kernel of $p_i$.  
Then  $M(S_i) \simeq M(S)/ker(p_i)$. 

\smallskip 
\noindent If $S_i \hookrightarrow S_j$, 
then we denote the canonical projection by 
$p^j_i : M(S_j) \rightarrow M(S_i)$, and 
if  $S_i, S_j \in {\cal S}$, then 
$p^j_{ij} : M(S_j) \rightarrow M(S_i \cap S_j)$, and 
$p^i_{ij} : M(S_i) \rightarrow M(S_i \cap S_j)$ 
are the canonical mappings.
Note that 
all homomorphisms $p^i_j : M(S_i) \rightarrow M(S_j)$ and 
$p^i_{ij} : M(S_i) \rightarrow M(S_i \cap S_j)$ are onto.
We know that 
for all $S_j \hookrightarrow S_i$, $p^i_j \circ p_i = p_j$.

\begin{example}
\label{ex-trains-trace}
Consider a family of two systems of trains $S_1, S_2$ over 
disjoint sets $I_1, I_2$ of trains as in 
Example~\ref{interconnect-3sys} but with $l < k$.
We simplify the description by 
replacing all actions that need to be executed at the 
same time with one action.
The system $S_i$ ($i \in \{ 1, 2 \}$) 
obtained this way has two actions ${\sf update}_i$
and ${\sf move}_i$  
The constraints 
are $C_i = \{ {\sf update}_i  \wedge {\sf move}_i = 0\}$. 
Thus $\theta_i = id$, so $M(S_i) = A_{S_i}^*$.

\noindent Let $S$ be the system obtained by the interconnection of 
$S_1$ and $S_2$. 

\medskip
\noindent $A_S= \{ {\sf update}_1, {\sf update}_2, 
{\sf move}_1, {\sf move}_2 \}$ and $C_S = C_1 \cup C_2$.

\medskip
\noindent $\begin{array}{@{}ll}
D_S = & \{ ({\sf update}_1, {\sf update}_1), ({\sf update}_2, {\sf update}_2),
({\sf move}_2, {\sf move}_2), ({\sf move}_1, {\sf move}_1), \\
& ~   ({\sf update}_1, {\sf move}_1),({\sf move}_1, {\sf update}_1), ({\sf update}_2, {\sf move}_2),({\sf move}_2, {\sf update}_2) \} \\
\end{array}$

\medskip
\noindent $\begin{array}{@{}rl}
(A_S \times A_S) \backslash D_S = & \{ ({\sf update}_1, {\sf update}_2),({\sf update}_2, {\sf update}_1),({\sf update}_1, {\sf move}_2),\\
& ~({\sf move}_2, {\sf update}_1),({\sf move}_1, {\sf update}_2),({\sf update}_2, {\sf move}_1),\\
& ~({\sf move}_1, {\sf move}_2), ({\sf move}_2, {\sf move}_1) \} \\
\end{array}$

\medskip
\noindent Thus, $M(S) = A_S^*/\theta$, where $\theta$ is the congruence generated by 
$(A_S \times A_S) \backslash D_S$. 

\end{example}

\medskip
\noindent 
Applying a method due to \cite{Davey} (cf.\ Appendix~\ref{app:davey}) 
-- where sheaves of algebras are constructed, 
whose stalks are
quotients of a given algebra -- 
we deduce for partially commutative monoids results similar to those 
given in \cite{MonteiroPereira86} for monoids. The results are 
similar to results on Petri Nets and Mazurkiewicz traces presented in 
\cite{Diekert90}.

\medskip
\noindent
Let $(F, f, {\sf InSys})$ be defined by 
$F = \coprod_{S_i \in {\sf InSys}} M(S_i)$, and 
$f : F \rightarrow {\sf InSys}$ be the natural projection. 
Assume that a subbasis for the topology on $F$ is 
${\cal S}{\cal B} = \{ [m](U) \mid U \in \Omega({\sf InSys}), m \in M(S) \}$, where 
$[m](U) = \{ p_i(m) \mid i \in U \}$. 
 
\medskip
\noindent We first show that 
$\Omega({\sf InSys})$ has the property that 
for every $m_1, m_2 \in M(S)$, if 
$p_i(m_1) = p_i(m_2)$ then there exists an open neighborhood $U$ 
of $S_i$ in ${\sf \Omega(InSys)}$ such that for every 
$S_j \in U$, $p_j(m_1) = p_j(m_2)$ (i.e.\ 
it is an S-topology). 

\begin{lemma}
$\Omega({\sf InSys})$ is a S-topology (cf.\  
Definition~\ref{s-topology}). 
\end{lemma}
\Proof We show that for every $m_1, m_2 \in M(S)$, if 
$p_i(m_1) = p_i(m_2)$ then there exists an open neighborhood $U$ 
of $S_i$ in ${\sf \Omega(InSys)}$ s.t.\ for every 
$S_j \in U$, $p_j(m_1) = p_j(m_2)$.  
Let $m_1, m_2 \in M(S)$ with $p_i(m_1) = p_i(m_2)$. 
Let $U = {\downarrow} S_i = \{ S_j \in {\sf InSys} \mid S_j \hookrightarrow S_i \}$. $U \in \Omega({\sf InSys})$ and  $p_j(m_1) = p^i_j(p_i(m_1)) = p^i_j(p_i(m_2)) = 
p_j(m_2)$ 
for every $S_j \in U$. \QED
 
\medskip
\noindent 
Let $\alpha : M(S) \rightarrow \Gamma(I, F_A)$ be defined by
 $\alpha(m) = ([m]_{\theta_i})_{i \in I}$. 
Since $\Omega({\sf InSys})$ is an S-topology, 
by Theorem~\ref{Davey} and Corollary~\ref{CorDavey} 
 in Appendix~\ref{app:davey} we have:

\begin{quote}
\medskip
 (1) $(F, f, {\sf InSys})$ is a sheaf of algebras,
 
\smallskip
 \noindent
(2) The stalk at $S_i \in {\sf InSys}$  is isomorphic to $M(S_i)$, 
 
\smallskip
\noindent (3)  In 
 $M(S) \stackrel{\alpha}{\rightarrow} \Gamma({\sf InSys}, F) \leq 
 \prod_{S_i \in {\sf InSys}} M(S_i) \stackrel{\pi_i}{\rightarrow} M(S_i)$
 
~(3.i) $\pi_i \circ \alpha$ is an epimorphism, 

~(3.ii) $M(S)$  is a subdirect
 product of  $\{ M(S_i) \}_{S_i \in {\sf InSys}}$
  iff $\alpha$ is a monomorphism. 
  \end{quote}

\begin{lemma} 
Let $s :  {\sf InSys} \rightarrow 
\coprod_{S_i \in {\sf InSys}} M(S_i)$ be such that 
$s(S_i) \in M(S_i)$ for every $S_i \in {\sf InSys}$. Let 
$m \in M(S)$ and $U \in \Omega({\sf InSys})$. Then 
$S_i \in s^{-1}([m](U))$ if and only if 
$S_i \in U$ and $s(S_i) = p_i(m)$.
\label{s-1}
\end{lemma}

\Proof 
Note that 
$s^{-1}([m](U)) = \{ S_i \in {\sf InSys} \mid s(S_i) \in [m](U) \} = 
\{ S_i \in {\sf InSys} \mid s(S_i) \in \{ p_j(m) \mid S_j \in U \} \}$. 
We first prove the direct implication.  
Assume that $S_i \in s^{-1}([m](U))$. Then $s(S_i) = p_j(m)$ for some 
$S_j \in U$. Since $f \circ s(S_i) = S_i$, it follows that 
$S_i = f(s(S_i)) = f(p_j(m)) = S_j$, hence $S_i \in U$ and $s(S_i) = p_i(m)$.
To prove the converse, 
assume that $S_i \in U$ and $s(S_i) = p_i(m)$. Then 
$s(S_i)  \in \{ p_j(m) \mid S_j \in U \}$, hence $S_i \in s^{-1}([m](U))$.
\QED

\medskip

\begin{lemma} 
Let $\tau$ be the topology on $F = \coprod_{S_i \in {\sf InSys}} M(S_i)$
generated by ${\cal SB}  = \{ [m](U) \mid U \in \Omega({\sf InSys}), m \in M(S) \}$ as a subbasis. 
Then any map $$s :  {\sf InSys} \rightarrow 
\coprod_{S_i \in {\sf InSys}} M(S_i)$$ such that for every 
$S_i \in {\sf InSys}$, $s(S_i) \in M(S_i)$
is continuous if and only if 
for every $S_i, S_j \in {\sf InSys}$ such that 
$S_j \hookrightarrow S_i$, $p^i_j(s(S_i)) = s(S_j)$.
\label{s-cont}
\end{lemma}
\Proof 
Since ${\cal SB}$ is a subbasis for the topology on 
$F = \coprod_{S_i \in {\sf InSys}} M(S_i)$, a map 
$$s :  {\sf InSys} \rightarrow 
\coprod_{S_i \in {\sf InSys}} M(S_i)$$ is continuous iff 
for every $[m](U) \in {\cal SB}$, $s^{-1}([m](U)) \in \Omega({\sf InSys})$.
We first prove the direct implication. 
Assume that $s :  {\sf InSys} \rightarrow 
\coprod_{S_i \in {\sf InSys}} M(S_i)$ is continuous. 
Let $S_i, S_j \in {\sf InSys}$ be such that $S_j \hookrightarrow S_i$. 
We prove that $p^i_j(s(S_i)) = s(S_j)$. 
Let $U =~{\downarrow} S_i \in \Omega({\sf InSys})$ and let $m \in M(S)$ 
be such that $p_i(m) = s(S_i)$ (the existence of $m$ is ensured by the 
fact that $p_i : M(S) \rightarrow M(S_i)$ is onto). 
From the continuity of $s$ we know that 
$s^{-1}([m]({\downarrow} S_i)) \in \Omega({\sf InSys})$. Obviously,
$S_i \in s^{-1}([m]({\downarrow} S_i))$. Therefore, since $S_j \hookrightarrow 
S_i$, 
$S_j \in s^{-1}([m]({\downarrow} S_i))$, hence, by Lemma~\ref{s-1},
$s(S_j) = p_j(m)$. 
Therefore, $s(S_j) = p_j(m) = p^i_j(p_i(m)) = p^i_j(s(S_i))$.
 
Conversely, assume that for every $S_i, S_j \in {\sf InSys}$ such that 
$S_j \hookrightarrow S_i$ it holds that $p^i_j(s(S_i)) = s(S_j)$.
We prove that $s$ is continuous. 
Let $[m](U) \in {\cal SB}$, where 
$m \in M(S)$ and $U \in \Omega({\sf InSys})$.
We prove that $ s^{-1}([m](U)) \in \Omega({\sf InSys})$.
Let $S_i \in s^{-1}([m](U))$. Then  $S_i \in U$ and $s(S_i) = p_i(m)$.
Let $S_j \hookrightarrow S_i$. Then $S_j \in U$ and by the hypothesis, 
$s(S_j) = p^i_j(s(S_i)) = p^i_j(p_i(m)) = p_j(m)$. Thus, $S_j \in 
 s^{-1}([m](U))$. Therefore $ s^{-1}([m](U)) \in \Omega({\sf InSys})$. \QED

\medskip
\begin{lemma} The set $\Gamma({\sf InSys}, F)$ of global sections of $F$  
has the form 

\noindent 
$\Gamma({\sf InSys}, F) = \{ (m_i)_{S_i \in {\sf InSys}} \mid 
m_i \in M(S_i) \mbox{ and } \forall 
S_j \hookrightarrow S_i \in {\sf InSys}, p^i_j(m_i) = m_j \}.$

\label{global-sections}
\end{lemma} 
\Proof 
We know that $\Gamma({\sf InSys}, F) = \{ s : {\sf InSys} \rightarrow 
\coprod_{S_i \in {\sf InSys}} M(S_i) \mid s$  continuous and  
$s(S_i) \in M(S_i),  \forall S_i \in {\sf InSys} \}$. 
(The elements of $\Gamma({\sf InSys}, F)$ are tuples  
$(s(S_i))_{S_i \in {\sf InSys}}$.)  
Let first $s \in \Gamma({\sf InSys}, F)$. 
Then $s$ is continuous and, by Lemma~\ref{s-cont},  
for all $S_i, S_j \in {\sf InSys}$ with 
$S_j \hookrightarrow S_i$, 
$p^i_j(s(S_i)) = s(S_j)$. 
Conversely, let $(m_i)_{S_i \in {\sf InSys}}$ be such that for every 
$S_i, S_j \in {\sf InSys}, m_i \in M(S_i)$ if 
$S_j \hookrightarrow S_i$ then $ p^i_j(m_i) = m_j$.
Let  $s : {\sf InSys} \rightarrow \coprod_{S_i \in {\sf InSys}} M(S_i)$ be 
defined by $s(S_i) = m_i$ for every $S_i \in {\sf InSys}$. 
Then, whenever $S_j \hookrightarrow S_i \in {\sf InSys}$, 
$p^i_j(s(S_i)) = s(S_j)$ and, by Lemma~\ref{s-cont}, 
$s$ is continuous. \QED

\medskip
\begin{theorem}
Let $(F, f, {\sf InSys})$ be defined as above. Then 
$(F, f, {\sf InSys})$ is a sheaf space of algebras. 
The stalk at $S_i \in {\sf InSys}$  is isomorphic to $M(S_i)$; 
the set of global sections is 
$$\Gamma({\sf InSys}, F) = \{ (m_i)_{S_i \in {\sf InSys}} \mid 
m_i \in M(S_i), \mbox{ and } \forall 
S_i \hookrightarrow S_j, p^j_i(m_j) = m_i \}.$$
Additionally the following hold: 
\begin{enumerate}
\item[(1)] If {\sf InSys} is finite, then 
\begin{enumerate}
\item[(i)] $M(S) \hookrightarrow  \Gamma({\sf InSys}, F) \leq 
\prod_{S_i \in {\sf InSys}} M(S_i) \stackrel{\pi_i}{\rightarrow} M(S_i)$
is a subdirect product. 

\item[(ii)] The embedding $M(S) \hookrightarrow  \Gamma({\sf InSys}, F)$ is an 
isomorphism iff every chordless cycle in the 
dependence graph $G_S$ of $S$ is a cycle in a subgraph $G_{S_i}$ for some 
$S_i \in {\sf InSys}$. 
\end{enumerate}
\item[(2)] If {\sf InSys} is infinite,  and if 
for every $a \in A_S$  there are at most finitely many 
$S_i \in {\sf InSys}$ with
$a \in A_i$,  then 
there is an injective morphism $M(S) \rightarrow \bigoplus_{S_i} M(S_i)$,
where $\bigoplus_{S_i} M(S_i) = \{ (w_i)_{i \in I}  \mid  w_i \in M(S_i), 
w_i = \varepsilon \mbox{ a.e.} \}$ is the weak product of 
the family $\{ M(S_i) \}_{S_i \in \mbox{ {\sf InSys}}}$.
\end{enumerate}
\label{sheaves-pcmon}
\end{theorem}
\Proof The form of $\Gamma({\sf InSys}, F)$
follows  from Lemma~\ref{global-sections}. 
(1)(i) and (2) are a consequence of Theorem~\ref{GET} and the subsequent 
comments in Appendix~\ref{app:pcm}. (1)(ii) is a direct consequence of Theorem~3.3.2 in \cite{Diekert90}. \QED

\begin{example}
First consider the family of systems in 
Example~\ref{ex-trains-trace}. The dependency graph of $S$,  
$G_S = (A_S, D_S)$ contains the following non-trivial 
chordless cycles:  
\begin{enumerate}
\item $( {\sf update}_1, {\sf move}_1, {\sf update}_1 )$ and  
$( {\sf move}_1, {\sf update}_1, {\sf move}_1 )$ (all cycles in $G_{S_1}$) 
\item $( {\sf update}_2, {\sf move}_2, {\sf update}_2 )$ and 
$( {\sf move}_2, {\sf update}_2, {\sf move}_2 )$ (all cycles in $G_{S_2}$).
\end{enumerate}
Thus, in this case the embedding in Theorem~\ref{sheaves-pcmon}(1)(ii)
is an isomorphism. 
\end{example}
\begin{example}
Consider the systems in Example~\ref{ex-onto}. 
The dependency graphs are: 
\begin{itemize}
\item $G_{S_1} = (A_1, D_1)$, with $D_1 = \{ (a, a), (b, b), (d, d), (a, b),(b, a) \}$, 
\item $G_{S_2} = (A_2, D_2)$, with $D_2 = \{ (b, b), (c, c), (e, e), (b, c), (c, b) \}$, 
\item $G_{S_3} = (A_3, D_3)$, with $D_3 = \{ (a, a), (c, c), (f, f), (a, c), (c, a) \})$. 
\end{itemize} 
$G_S = (A_1 \cup A_2 \cup A_3, D_1 \cup D_2 \cup D_3)$ contains the chordless 
cycle $(a, b, c, a)$ which is not contained in any of the subgraphs $G_{S_i}, i \in \{ 1, 2, 3 \}$. Thus, the embedding in Theorem~\ref{sheaves-pcmon}(1)(ii)
is not an isomorphism. 
\end{example}

\section{Other concepts and their sheaf semantics}

\noindent 
{\bf Time.} One possibility for  expressing  time internally in the category 
${\sf Sh}({\sf InSys})$ is to model time 
by the sheafification ${\mathbb N}$ of the constant presheaf 
${\cal N} : \Omega({\sf InSys})^{\mbox{\scriptsize op}} \rightarrow {\sf Set}$
(defined for every $U$ by ${\cal N}(U) = {\mathbb N}$), 
which can be constructed as follows: 
\begin{itemize}
\item Let  
${\cal N}^+ : \Omega({\sf InSys})^{op} \rightarrow {\sf Sets}$, 
defined by 
${\cal N}^+(U) = {\Bbb N}$ if $U \neq \emptyset$ and 
${\cal N}^+(\emptyset) = 1$ (for the empty cover there is exactly one 
matching family; the empty one). 

\item Let  
${\mathbb N} = ({\cal N}^+)^+  : 
\Omega({\sf InSys})^{op} \rightarrow {\sf Sets}$. 
An element of $({\cal N}^+)^+(U)$ is an equivalence class of sets of elements
$i_j \in {\cal N}(U_j)$ for some open covering 
$\{ U_j \mid j \in J \}$ of $U$, 
which match ($i_{j_1} = i_{j_2}$) whenever the overlap 
$U_{j_1} \cap U_{j_2}$ is nonempty. 
Thus, these elements ``glue'' together to give a function 
$i : U \rightarrow {\Bbb N}$, with the property that every point of 
$U$ has some open neighborhood on which the function is constant. 
\end{itemize}
For every $U \in \Omega({\sf InSys})$, 
${\Bbb N}(U) = 
\{ i : U \rightarrow {\Bbb N} \mid f \mbox{ locally constant}\footnote{ 
$f {:} U {\rightarrow} X$ is locally constant if 
$\forall x {\in} U$ there is an open neighborhood $U_1 {\subseteq} U$ of $x$
on which $f$ is constant.
This means that
 'local clocks' of the systems in $U$ synchronize 
for systems sharing common subsystems.}\}$. There exist {\sf Sh(InSys)}-arrows  
$1 \stackrel{0}{\rightarrow}{\Bbb N} \stackrel{s}{\rightarrow}{\Bbb N};$ 
the sheaf ${\Bbb N}$ is the natural number object in  {\sf Sh(InSys)}.

\medskip
\noindent 
{\bf Other constructions.} 
Various other sheaves and natural transformations can be defined by using 
standard categorical constructions in ${\sf Sh}({\sf InSys})$. 
We can e.g.\ define a natural transformation 
$\mbox{{\sf B}}_{\mathbb N} \times {\mathbb N} 
\stackrel{{\sf a}}{\rightarrow} \mbox{\sf St} \times \mbox{\sf Pa}$ 
whose components 
$\mbox{\sf B}_{\mathbb N}(U) 
\times {\mathbb N}(U) \stackrel{{\sf a}_U}{\rightarrow} 
\mbox{\sf St}(U) \times \mbox{\sf Pa}(U)$ 
are defined by 
${\sf a}_U(h, (n_i)_{S_i \in U}) = ((s^i_i)_{S_i \in U}, (f^i_i)_{S_i \in U})$,
for every $U \in \Omega({\sf InSys})$, 
where for every $S_i \in U$, 
$h(n_i) = ((s^i_j)_{S_j \in U}, (f^i_j)_{S_j \in U})$. 
\footnote{
The map ${\sf a}_U$ has as arguments a behaviour along 
${\mathbb N}$ of the 
family of systems in $U$, $h \in \mbox{\sf B}_{\mathbb N}(U)$, and a
tuple consisting of 'local clocks' of the systems in $U$ which synchronize 
on systems sharing common subsystems. 
${\sf a}_U$ returns the pair 
$((s^i_i)_{S_i \in U}, (f^i_i)_{S_i \in U})$ where 
$(s^i_i, f^i_i)$ is the pair state/parallel action in the behavior 
corresponding to the system $S_i$ in $U$, 
at the time point indicated by the local clock $n_i$ of $S_i$.}

\begin{theorem}[\cite{Sofronie-Stokkermans99}]
For every $S_i \in {\sf InSys}$, ${\sf Stalk}_{S_i}({\sf a})$
is (up to isomorphism)  the map
$B_T(S_i) \times {\mathbb N} \stackrel{{\sf a}_{S_i}}{\rightarrow} St(S_i)
\times Pa(S_i)$,
defined by ${\sf a}_{S_i}(h, n) = h(n)$.
\label{prop3}
\end{theorem}

\vspace{-2mm}
\section{Geometric logic and properties of systems}
\label{geo}

\vspace{-2mm}
We provide interpretations for
properties of systems (i.e.\ statements about 
states, actions, behavior) both concretely (in the category of sets) 
and in a category of sheaves, and establish links between the 
set-theoretical (both for individual systems and for their interconnections) and the sheaf-theoretical interpretation. These
links are then used to prove preservation of truth when interconnecting 
systems.

\vspace{-2mm}
\subsection{Many-sorted first order languages and their interpretation in 
$\mbox{\sf Sh}(I)$} 
Let ${\cal L}$ be a many-sorted  first-order 
language consisting of a collection of sorts and  collections of function 
and relation symbols. 
Terms and atomic formulae from ${\cal L}$ are defined 
in the standard way; compound formulae are constructed by
using the connectives 
$\vee, \wedge, \Rightarrow, \neg$ and the quantifiers $\exists, \forall$, 
for every sort $X$.
An {\em interpretation $M$ of ${\cal L}$ in $\mbox{\sf Sh}(I)$} is constructed
by associating: 
\begin{itemize}
\item a sheaf $X^M$ on $I$ to every sort $X$,
\item a subsheaf $R^M \subseteq X_1^M \times \dots  \times X_n^M$ to
every relation symbol $R$ of arity $X_1 \times \dots  \times X_n$, 
\item an arrow $f^M : X_1^M \times \dots  \times X_n^M \rightarrow Y^M$
in $\mbox{\sf Sh}(I)$ to every function symbol 
$f$ with arity $X_1 \times\dots\times X_n \rightarrow  Y$.
\end{itemize}
Each term $t(x_1, \dots, x_n)$ of sort $Y$ is (inductively) 
interpreted as an arrow
$t^M : X_1^M \times \dots  \times X_n^M \rightarrow Y^M$; and 
every formula $\phi(x_1, \dots, x_n)$ with free variables
$FV(\phi) \subseteq \{ x_1, \dots, x_n \}$, where $x_i$ is of sort $X_i$, 
gives rise to 
a subsheaf $\{ (x_1, \dots, x_n)  \mid  \phi(x_1, \dots, x_n) \}^M \subseteq 
X_1^M \times \dots  \times X_n^M$. 
For details we refer to \cite{MacLaneMoerdijk92}, Ch.~X.
\begin{definition}
A {\em geometric formula} is a formula built from atomic
formulae by using only the connectives
$\vee$ and  $\wedge$  and the quantifier $\exists$. 
A {\em geometric axiom} 
is a formula of the form $(\forall x_1, \dots, x_n)(\phi \Rightarrow \psi)$
where $\phi$ and $\psi$ are geometric formulae.

Let ${\mathbb T}$ be a theory in the language ${\cal L}$.
A variable in a geometric formula 
is called {\em ${\mathbb T}$-provably unique} if its value
in every model of ${\mathbb T}$ is uniquely determined by the values
of the remaining free variables.

A {\em cartesian formula w.r.t.\ ${\mathbb T}$} 
is a formula constructed from atomic
formulae using only the connective
$\wedge$ and the quantifier $\exists$ 
over ${\mathbb T}$-pro\-va\-bly unique variables.
A {\em car\-te\-si\-an axiom w.r.t.\ ${\mathbb T}$} 
is a formula of the form $(\forall x)(\phi(x) \Rightarrow \psi(x))$
where $\phi$ and $\psi$ are cartesian formulae w.r.t.\ ${\mathbb T}$.
A {\em cartesian theory} is a theory whose axioms can be ordered such that
each is cartesian w.r.t.\ the preceding ones.
\end{definition}

\noindent A geometric axiom
$(\forall x_1 \dots x_n)(\phi {\Rightarrow} \psi)$
is {\em satisfied in an interpreta\-tion $M$  in $\mbox{\sf Sh}(I)$}
if ${\{} (x_1,  \dots, x_n)  {\mid}  \phi {\}}^M$ is a subobject of
${\{} (x_1,  \dots, x_n)  {\mid}  \psi {\}}^M$ in $\mbox{\sf Sh}(I)$.

\vspace{-2mm}
\subsection{Stalk functors, global section functors; preservation of truth}

\noindent {\bf Stalk functors.}
For every $S_i \in {\sf InSys}$ let $f_i : \{ * \} \rightarrow {\sf InSys}$
be defined by $f_i (*) = S_i$.
The inverse image functor corresponding to $f_i$, the stalk 
functor 
${\sf Stalk}_{S_i} = f_i^* : {\sf Sh}({\sf InSys}) \rightarrow 
{\sf Set}$, associates to every sheaf $F \in  \mbox{\sf Sh}({\sf InSys})$ the 
stalk at $S_i$, $F_{S_i}$.
For all $S_i \in {\sf InSys}$, 
$f_i^*$ preserves the validity of geometric axioms. 
The stalk functors $ f_i^*$ 
are collectively faithful, so they reflect the validity of 
geometric axioms. 

\medskip
\noindent {\bf Global section functor.}
Consider the unique map $g : \mbox{{\sf InSys}} \rightarrow \{ * \}$.
The direct image functor, 
$g_* : {\sf Sh}({\sf InSys}) \rightarrow {\sf Set}$, is the
global section functor
$g_*(F) = F(\mbox{{\sf InSys}})$ for every $F \in 
{\sf Sh}({\sf InSys})$. Thus, the global section functor
preserves the interpretation of every cartesian axiom.

\vspace{-2mm}
\subsection{A geometric logic for reasoning about complex systems} 
Let ${\cal L}$ be a fixed many-sorted language including at least
sorts like 
{\sf st}(ate), 
{\sf pa}(rallel-action), 
{\sf b}(ehavior), 
{\sf t}(ime); 
constants like 
$s_0 : {\sf st}$ (initial state), 
$0 : {\sf t}$ (initial moment of time);
function symbols like 
\begin{itemize}
\item 
${\sf appl} : {\sf b} \times {\sf t} \rightarrow {\sf st} \times {\sf pa}$, 
\item ${\sf p_1} : {\sf st} \times {\sf pa} \rightarrow {\sf st}$, 
\item ${\sf p_2} : {\sf st} \times {\sf pa} \rightarrow {\sf pa}$; 
\end{itemize}
relation symbols like 
\begin{itemize}
\item {\sf tr}(ansition) $\subseteq$ {\sf pa} $\times$ {\sf st} $\times$ {\sf st}, 
\item $=_X \subseteq X \times X$ for every sort $X$, etc.
\end{itemize}
Let $M$ be an interpretation of ${\cal L}$ in ${\sf Sh}({\sf InSys})$
such that 
\begin{itemize}
\item {\sf st}$^M = {\sf St}$, 
{\sf pa}$^M = {\sf Pa}$, 
{\sf b}$^M = {\sf B}_{\mathbb N}$, ${\sf t} = {\mathbb N}$, 
${\sf appl}^M = {\sf a}$, 
\item ${\sf p_1}^M = \pi_1,
{\sf p_2}^M = \pi_2$ (the canonical projections),  
\item ${\sf tr}^M = {\sf Tr}$. 
\end{itemize}
For every sort $X$, we interpret 
$=_{X} : X \times X \rightarrow \Omega $ 
as usual.

\begin{theorem}[\cite{Sofronie-Stokkermans99}]
${\sf Sh}({\sf InSys})$ satisfies a geometric axiom 
in the interpretation $M$ if and only if 
\, {\sf Set} satisfies 
it in all interpretations $f_i^*(M)$.
If ${\sf Sh}({\sf InSys})$ satisfies a cartesian axiom, 
this is also true in {\sf Set} in the interpretation 
$g_*(M)$ ($f_i^*(M)$ and $g_*(M)$ interpret a sort $X$ as 
$f_i^*(X^M)$ resp.\ $g_*(X^M)$). 
\end{theorem}

\noindent 
From Theorems~\ref{thm1}~and~\ref{thm3} we know that 
for every $S_i \in {\sf InSys}$, 
$$f^*_i({\sf St}) = {\sf St}_{S_i} \simeq St(S_i) \text{ and } f^*_i({\sf Pa}) = {\sf Pa}_{S_i} \simeq Pa(S_i);$$
if $S$ is the system obtained by interconnecting all elements in
{\sf InSys}, 
$$g_*({\sf St}) = {\sf St}({\sf InSys}) \simeq St(S) \text{ and } g_*({\sf Pa}) = {\sf Pa}({\sf InSys}) \simeq Pa(S).$$
The same holds for  ${\sf Tr}$ and ${\sf B}_T$.  
Moreover, $f^*_i({\mathbb N}) = {\mathbb N}$, 
$g^*({\mathbb N}) = {\mathbb N}({\sf InSys})$, and, by Theorem~\ref{prop3},  
$$f^*_i({\sf appl}) = {\sf a}_{S_i} : B_{\mathbb N}(S_i) \times {\mathbb N} \rightarrow St(S_i) \times Pa(S_i).$$
Hence, statements about states, 
actions and
transitions in ${\sf Sh}({\sf InSys})$ are translated by
$f_i^*$ (resp.\ $g_*$) to corresponding statements
about states, actions and transitions in $S_i$ (resp.\ $S$).

\medskip
\noindent 
We illustrate the ideas above by several classes of 
properties of systems (adapted from \cite{Kroger87})
which we express in the language ${\cal L}$. For instance, 
if $h$ is a possible behavior and 
$j$ a moment in time, then $h(j)$ 
can be expressed in ${\cal L}$ by ${\sf appl}(h, j)$;
the state of $h$ at $j$ can be expressed by ${\sf s}(h, j)$, where
$${\sf s} = {\sf p_1} \circ {\sf appl}  
: {\sf b} \times {\bf t} \stackrel{\sf appl}{\longrightarrow} {\sf st} \times {\sf pa} \stackrel{\sf p_1}{\longrightarrow} {\sf st}.$$

\begin{description}
\item[(a) Safety properties]
are of the form
$$(\forall h: {\sf b})(\forall j : {\sf t}) 
( P({\sf s}(h,0)) \Rightarrow Q({\sf s}(h,j)) ),$$
where $P$ and $Q$ are formulae in ${\cal L}$. As examples we mention:

\medskip 
\noindent {\em ~(i) Partial correctness:}
$$ (\forall h: {\sf b})(\forall j : {\sf t})[( P({\sf s}(h, 0)) \wedge 
{\sf Final}({\sf s}(h, j))) \Rightarrow Q({\sf s}(h,j))]; $$ 
\noindent {\em (ii) Global invariance of $Q$:}
$$ (\forall h: {\sf b})(\forall j : {\sf t})[ P({\sf s}(h,0)) 
\Rightarrow Q({\sf s}(h,j))]. $$
\item[(b) Liveness properties] have the form
$$(\forall h: {\sf b})[P({\sf s}(h,0)) \Rightarrow (\exists j : {\sf t})Q({\sf s}(h,j))].$$
With $s_0$ denoting the initial and $s_f$ a final state,
examples are:

\medskip
\noindent {\em ~(i) Total correctness and termination: }
$$(\forall h: {\sf b})[ P({\sf s}(h,0)) \Rightarrow (\exists j : {\sf t})
({\sf Final}({\sf s}(h,j)) \wedge Q({\sf s}(h,j)))];$$
\noindent {\em (ii) Accessibility: }
$$(\forall h: {\sf b})[({\sf s}(h,0) = s_0) \Rightarrow (\exists j : {\sf t})({\sf s}(h,j) = s_f)].$$
\item[(c) Precedence properties:]
$$(\forall h: {\sf b})(\forall j : {\sf t}) [(P({\sf s}(h,0)) {\wedge} 
A({\sf s}(h,j))) {\Rightarrow} Q({\sf s}(h,j))].$$
\end{description}

\begin{theorem}[\cite{Sofronie-Stokkermans99}]
Assume that the following conditions are fulfilled: 
\begin{enumerate}
\item[(1)]  The final states
form a subsheaf
${\sf St}_f \subseteq {\sf St}$ interpreting a sort ${\sf st_f}$ of ${\cal L}$.
(This happens
e.g.\ if in the definition of a system
final states are specified by additional constraints, 
and in defining colimits this information is also used.)

\medskip
\item[(2)]   The properties $P, Q, A$ can be expressed in 
${\cal L}$ (using the sorts,  
constants, function and relation symbols mentioned at the beginning
of Section~\ref{geo}), 
and can be interpreted in ${\sf Sh}({\sf InSys})$
and also in {\sf Set} (to express, for every $S_i$ in
{\sf InSys}, the corresponding property of $S_i$, or $S$).
\end{enumerate}
 The truth of formulae describing safety, 
liveness and precedence properties (as in 
(a),(b),(c) above) is 
preserved under inverse image functors if in the definitions of
the property $P$ (c.q.\ $Q, A$) only conjunction, disjunction
and existential quantification occur.  The truth of these formulae 
is additionally preserved
by direct image functors if only conjunction and unique existential
quantification occur in them.
\label{thm4}
\end{theorem}

\subsection{Example 1: Safety of train system controlled by radio 
controller}
Consider the example in 
Section~\ref{interconnect-3sys}: 
Let $k \leq l \in \{ 1, \dots, n \}$, 
$I_1 = \{ k, \dots, n \}, I_2 = \{ 1, \dots, l \}$, and 
$I_{12} = \{ k, \dots, l \}$. 
Let ${\sf InSys} = \{ S_1, S_2, S_{12}\}$ 
be the family consisting of the 
subsystems of $S$ described in Section~\ref{example-system}
corresponding to the sets of trains with indices in $I_1, I_2$ and $I_{12}$. 
Let $\Gamma_s^j$, $j \in \{ 1, 2, 12 \}$ be the following 
constraints encoding collision freeness of $S_j$ (where $\Rightarrow$ denotes logical implication):
$$\Gamma_s^j = \{ {\sf succ}({\sf TrainIndex}_i) = 
{\sf TrainIndex}_k \Rightarrow 
{\sf ActualPos}_i {<} {\sf ActualPos}_k {-} {\sf L} \mid i, k \in I_j \}.$$
For every $S_j \in \{ 1, 2, 12\}$ let 
${\sf SafeSt}(S_j) = \{ s : X_j \rightarrow M_j \mid s \models \Gamma_j \cup \Gamma_s^j \}$
be the set of safe states of $S_j$\footnote{We denote by $\Gamma_j$ the restriction of $\Gamma$ (cf.\ Definition~\ref{example-system}) to $X_j$}. 
Let 
$${\sf SafeState} : \Omega({\sf InSys})  \rightarrow {\sf Sets}$$ 
be defined: 
\begin{itemize}
\item on objects: by 
${\sf SafeState}(U) = \{ (s_j)_{S_j \in U} \mid s_j \in {\sf SafeSt}(S_j), 
\text{ and } {s_j}_{|X_i} = s_i \text{ whenever } S_i \hookrightarrow S_j \}$, 
and 
\item on morphisms: by restriction.
\end{itemize}
We can define a set of similar constraints $\Gamma_s$ and a similar 
set of safe states ${\sf SafeSt}(S)$ for the system $S$, where: 
$$\Gamma_s {=} \{ {\sf succ}({\sf TrainIndex}_i) {=}  
{\sf TrainIndex}_k \Rightarrow 
{\sf ActualPos}_i {<} {\sf ActualPos}_k {-} {\sf L} \mid 1 \leq i, k \leq n \}.$$
If $I_1 \cap I_2 \neq \emptyset$ then 
$\Gamma_s^1 \cup \Gamma_s^2 = \Gamma_s$\footnote{Note that if 
$I_1 \cap I_2 = \emptyset$ then some of the constraints of 
$\Gamma_s$ cannot be deduced from 
$\Gamma_s^1$ and $\Gamma_s^2$}. Analogously to Theorem~\ref{thm1} we can show: 
\begin{theorem} 
\label{thm-safe-state}
The following hold:
\begin{enumerate}
\item ${\sf SafeState}$ is a sheaf. Moreover,  ${\sf SafeState}$ is a 
subsheaf of ${\sf St}$. 

\smallskip
\item For each $S_i {\in} {\sf InSys}$, the stalk 
of ${\sf SafeState}$ at $S_i$  is in bijection with ${\sf SafeSt}(S_i)$.

\smallskip
\item ${\sf SafeState}({\sf InSys})$ is in bijection with ${\sf SafeSt}(S)$. 
\end{enumerate}
\label{sheaf-safe}
\end{theorem}

\noindent 
Collision freeness can be expressed as follows: 
$${\sf CollFree}~~~ (\forall h: {\sf b})(\forall j: {\sf t})~ [{\sf SafeState}({\sf s}(h, 0)) 
\Rightarrow {\sf SafeState}({\sf s}(h, j))].$$
This formula contains only atomic formulae and the implication 
symbol. Therefore, by Theorem~\ref{thm4}, its truth is preserved both 
under inverse image functors and under direct image functors, and it is 
reflected by the stalk functors: 
\begin{itemize}
\item Assume that $S_1, S_2, S_{12}$ satisfy ${\sf CollFree}$. 
Then for all $h \in B_{\mathbb N}(S_j)$, $t \in {\mathbb N}$, 
if ${\pi}_1(h(0)) \in {\sf SafeSt}(S_j)$ then ${\pi}_1(h(t)) \in 
{\sf SafeSt}(S_j)$. Due to the form of the formula ${\sf CollFree}$, 
its truth is reflected by the stalk functors 
$f^*_j : {\sf Sh}({\sf InSys}) \rightarrow {\sf Set}$. 
It therefore follows that ${\sf Sh}({\sf InSys})$ satisfies, 
internally, the formula  ${\sf CollFree}$. 

\smallskip
\item The truth of ${\sf CollFree}$ is preserved by the global section functor
$g_* : {\sf Sh}({\sf InSys}) \rightarrow {\sf Set}$, 
defined by $g(F) = F({\sf InSys})$. 
Therefore, (in {\sf Set}) the following holds: 
\end{itemize}
$$\begin{array}{ll}
\forall h \in  B_{\mathbb N}({\sf InSys}), \forall t \in  {\mathbb N}({\sf InSys}) &  [{\pi}_1(h(0)) \in {\sf SafeState}({\sf InSys}) \\ 
   & ~~ \Rightarrow  
{\pi}_1(h(t)) \in {\sf SafeState}({\sf InSys})]
\end{array}$$ 

\medskip
\noindent As, by Theorems~\ref{sheaf-safe} and~\ref{thm3}, 
${\sf SafeState}({\sf InSys})$ is in bijective correspondence with 
${\sf SafeSt}(S)$ and ${\sf B}_{\mathbb N}({\sf InSys})$ is in bijective 
correspondence with $B_{\mathbb N}(S)$, we obtain: 

\smallskip
\noindent $~ \forall h \in B_{\mathbb N}(S)$, $\forall t \in {\mathbb N}$, 
if ${\pi}_1(h(0)) \in {\sf SafeSt}(S)$ then 
${\pi}_1(h(t)) \in {\sf SafeSt}(S)$.

\begin{corollary}
Consider a family of consecutive trains on a linear track without loops.
Assume that each train $i$ controls both its  
position and the position of its predecessor, and accordingly 
determines its movement mode. We obtain a 
family $\{ S_i \mid i \in \{ 2, \dots, n \} \}$  of systems consisting  
of two successor trains each (each defined as in 
Example~\ref{example-system} for $n = 2$). 
Let $U$ consist of this family of systems 
together with their intersections. 
The colimit of this family is the system $S$ described in 
Example~\ref{ex-colimit}.  
By Theorem~\ref{thm4}, if collision freeness can be guaranteed for all the systems in $U$, 
then the system $S$ is collision free. 
\end{corollary}

\noindent For suitably chosen ${\sf minSpeed}, {\sf maxSpeed}$ and 
update interval $\Delta t$ all 2-train systems are collision free 
(for an automatic 
proof ideas from \cite{Jacobs-Sofronie-pdpar06} can be used).
Therefore, 
the $n$-train system in Example~\ref{example-system} can be proved to be 
collision free for these values.

\medskip
\noindent 
{\bf Remark:} The condition that the systems consist of successive trains and 
overlap over one extremity is needed for recovering the 
successor constraints on trains for the colimit. We obtain 
similar links between global and local properties also with 
a cover consisting of one-train systems. However, then  
the colimit of the system defined by such a cover is different 
of the system $S$; 
we would obtain a link between 
the safety of the systems consisting of one 
train only  and the safety of a system in which all trains are on independent 
tracks.

\subsection{Example 2: Lifeness}

We adapt the example in the previous section and give an 
example of lifeness property which can be expressed 
by means of a cartesian theory, and thus can be checked modularly.
Assume that the 
constraints $\Gamma'_j$ on for system $S_j$ 
consist of $\Gamma_j$ (defined as $\Gamma^l_k$ in 
Example~\ref{example-transition-connected}) and the constraint 
$(\bigwedge_{i \in I_j} {\sf Mode}_i = 0) \vee 
(\prod_{i \in I_j} {\sf Mode}_i > 0)$.
As in Theorem~\ref{thm-safe-state} 
we can prove that this defines a subsheaf 
${\sf St}'_j$ of ${\sf St}$; the following constraints define subsheaves of ${\sf St}'$ with properties similar to those of ${\sf SafeState}$: 
\begin{itemize}
\item $\Gamma^j_{su} = \Gamma'_j \cup \Gamma_s^j \cup \{ {\sf Mode}_i = 0 \mid i \in I_j \}$
defines a sheaf 
${\sf SafeStateUpdate}$; 
\vspace{-1mm}
\item $\Gamma^j_{\sf CanMove} =  \Gamma'_j \cup  \{ {\sf Mode}_i > 0 \mid i \in I_j \}$
defines a sheaf 
${\sf CanMove}$; 
\vspace{-1mm}
\item $\Gamma^j_{\sf CannotMove} =   \Gamma'_j \cup \{ {\sf Mode}_i = 0 \mid i \in I_j \}$
defines a sheaf 
${\sf CannotMove}$. 
\end{itemize}
For $S_i \in {\sf InSys}$ let 
${\sf Minimal}(S_i) = \{ (h, j) \mid s(h, j) \in {\sf CanMove}(S_i) 
\text{ and } 
\forall k ( s(h, k) \in {\sf CanMove}(S_i) \rightarrow k \geq j) \}$, 
characterizing the minimal moment in time $j$ w.r.t.\ a behavior $h$ at which 
all trains in system $S_i$ can move. 
These definitions can be used to define a subsheaf 
${\sf MinimalCanMove} \subseteq {\sf B}_{\mathbb N} \times {\mathbb N}$ 
with properties similar to those of ${\sf St}, {\sf Pa}, {\sf Tr}, {\sf B}$.
 A form of lifeness can be expressed by the following cartesian axioms:

\vspace{-1mm}
\begin{eqnarray*}
\forall h: {\sf b} & &  ({\sf SafeStateUpdate}({\sf s}(h,0)) \rightarrow \exists j : {\sf t} \,\, {\sf MinimalCanMove}(h, j)) \\
\forall h: {\sf b},  \forall i: {\sf t} & & ({\sf MinimalCanMove}(h, i) \rightarrow {\sf CanMove}({\sf s}(h, i)) )\\
\forall h: {\sf b},  \forall i, k: {\sf t} & & ({\sf MinimalCanMove}(h, i) \wedge {\sf CanMove}({\sf s}(h, k)) \rightarrow  i \leq k)
\end{eqnarray*}

\noindent (where the existential quantified variable in the first axiom is 
provably unique modulo the second and third axiom),  
and can thus be checked modularly.

\section{Conclusion}

We showed that a family {\sf InSys} of interacting systems
closed under pullbacks
can be endowed with a topology which models the way these systems interact.
States, parallel actions, transitions, and behavior can be described 
as sheaves on this topological space. 
We then used geometric logic 
to determine which kind of properties of systems in {\sf InSys} 
are preserved when interconnecting these systems. 
The main advantage of our approach is that it enables us 
to verify properties of complex systems in a modular way. 
We illustrated the ideas by means of a running example, involving 
systems of trains controlled by interacting controllers. In future 
work we plan to look at other applications, including geographically 
distributed systems, controlled by geographically fixed
controllers, whose domains overlap.  

\smallskip
\noindent We think that there should exist relationships between the 
approach described in this paper and other new approaches 
to the study of concurrency such as, for instance, higher dimensional 
automata (cf.\ \cite{Pratt1,Pratt2}) or approaches based on 
methods from geometry and algebraic topologicy in particular 
homotopic methods (cf.\ \cite{Herlihy95}).
Links between algebraic topology and concurrency as well as links with 
higher dimensional automata between have been studied e.g.\ by Gaucher, 
Goubault, Fajstrup, and Raussen 
(cf.\ e.g.\ \cite{Gaucher03,Fajstrup06}). We would like to compare our 
approach with the methods mentioned above. 
Using homological and especially homotopic 
methods seems to be the 
next natural step after the sheaf semantics given in this paper. 

\smallskip
\noindent {\bf Acknowledgements:} Many thanks to the referees for their helpful comments.

\medskip
\noindent This work was partly
  supported by the German Research
  Council (DFG) as part of the Transregional
  Collaborative Research Center ``Automatic
  Verification and Analysis of Complex
  Systems'' (SFB/TR 14 AVACS). See
  \texttt{www.avacs.org} for more information.

\newpage

\appendix
\section{Appendix. Sheaves of algebras}
\label{app:davey}

Let $A$ be an algebra of similarity type $\Sigma$, $(\theta_i)_{i
\in I}$ a family of congruences on $A$, and $\tau$ a
topology on $I$. The following problem was addressed and solved in
\cite{Davey}: In which situation does a sheaf exist with fibers $A_i
= A/\theta_i$ such that for every $a \in A$ the map $[a] : I
\rightarrow \coprod_{i \in I} A_i$ is a global section?
Two constructions are possible: 
\begin{description}
\item[{\bf Construction 1}]
Let $(F_A, f, I)$ be defined by $F_A =
\coprod_{i \in I} A/\theta_i$, and $f : F_A \rightarrow I$ be the
natural projection.  Assume that a subbasis for the topology on $F_A$
is $\{ [a](U) \mid U \in \tau, a \in A \}$, where $[a](U) = \{ [a](i)
\mid i \in U \} = \{ [a]_{\theta_i} \mid i \in U \} $.
\medskip
\item[{\bf Construction 2}] 
Let $G_A : \tau \rightarrow \Sigma Alg$ be
defined on objects by $G_A(U) = A/\theta_U$, where $\theta_U =
\bigwedge_{i \in U} \theta_i$ and on morphisms, for every $V \subseteq U$
by the canonical morphism $G_A(U) = A/\theta_U \rightarrow A/\theta_V
= G_A(V)$, $a_{\theta_U} \mapsto a_{\theta_V}$.

Let $G_i = \limright_{i \in U} G_A(U)$ be the stalks of $G_A$, 
and for every $i\in I$ let $g_i : G_i \rightarrow A_i$ be the unique 
morphism that arises from the universality property of the colimit.  
Note that $g_i(\rho^U_i(a)) = a_{\theta_i}$ for every $U \in \tau$ and 
every $i \in I$. $G_A$ is a presheaf of algebras. 
Let $(SG_A, g, I)$ be the associated sheaf.
\end{description}
In Construction 1, the stalk at $i$ is
isomorphic to $A_i$, but $(F_A, f, I)$ might be not a sheaf space.  In
Construction 2, $(SG_A, g, I)$ is a sheaf space, but
 $g_i : G_i \rightarrow A_i$ may not be an isomorphism. 

\vspace{-2mm}
\begin{theorem}[\cite{Davey}] The following conditions are
equivalent:
\begin{enumerate}
\item[(1)] If $[a]_{\theta_i} = [b]_{\theta_i}$ then there is
an open neighborhood $U$ of $i$ such that for every $j \in U$,
$[a]_{\theta_j} = [b]_{\theta_j}$.

\item[(2)] $(F_A, f, I)$ is a sheaf of
algebras.

\item[(3)] For every $i \in I$, $g_i : G_i \rightarrow A_i$ is
an isomorphism.  
\end{enumerate}
\label{Davey} 
\end{theorem}

\begin{definition} If $(\theta_i)_{i \in
I}$ is a family of congruences on an algebra $A$, then any topology on
$I$ that satisfies (1) is called an {\em S-topology}\index{S-topology}.
\label{s-topology}
\end{definition}
\begin{corollary}[\cite{Davey}] Assume that the topology on $I$ is
an S-topology with respect to the family of congruences $(\theta_i)_{i
\in I}$. Then $(F_A, f, I)$ and $(SG_A, g, I)$ are isomorphic sheaves
of algebras for which 
\begin{enumerate}
\item[(1)] The stalk at $i$ is isomorphic to $A_i = A/\theta_i$,

\item[(2)] The map $\alpha : A \rightarrow \Gamma(I, F_A)$ defined by
 $\alpha(a) = ([a]_{\theta_i})_{i \in I}$ is a homomorphism,

\item[(3)] In $A \stackrel{\alpha}{\rightarrow} \Gamma(I, F_A) 
\leq \prod_{i \in I} A/{\theta}_i \stackrel{p_i}{\rightarrow} A/{\theta}_i $:
\begin{quote} 
(i) $p_i \circ \alpha$ is an epimorphism, and \\  
(ii) $A$ is a subdirect
product of the family $( A/{\theta}_i )_{i \in I}$ iff
$\bigwedge_{i \in I} {\theta}_i = \Delta_A$ \\
$\mbox{\hspace{5mm}}$ (i.e.\ iff $\alpha$ is a monomorphism).  
\end{quote}
\end{enumerate}
\label{CorDavey} 
\end{corollary}

\noindent 
The coarsest S-topology on $I$ can be constructed as follows:
\begin{lemma}[\cite{Davey}, \cite{Krauss-Clark-1979}] 
Let $A \hookrightarrow \prod_{i \in I} A_i
\stackrel{p_i}{\rightarrow} A_i$ be a subdirect product. The coarsest
S-topology on $I$ is generated by the sets 
$E(a, b) {=} \{i {\in} I \mid p_i(a) {=} p_i(b) \}$ as a subbasis. 
\end{lemma}
\begin{lemma}[\cite{Krauss-Clark-1979}] Let $A \hookrightarrow
\prod_{i \in I} A_i \stackrel{p_i}{\rightarrow} A_i$ be a subdirect
product and ${\tau}_1, {\tau_2}$ be two topologies on $I$. If
${\tau}_1 \subseteq {\tau}_2$ and ${\tau}_1$ contains the equalizer
topology induced by $A$ (generated by the sets 
$E(a, b)$ as a subbasis), then $\Gamma(F_A, (I, \tau_1)) \subseteq
\Gamma(F_A, (I, \tau_2))$.  \end{lemma}
\noindent 
Even if the topology on $I$ is an S-topology, 
$A$ is not necessarily isomorphic to 
the algebra $\Gamma(I, F_A)$. 
A necessary and sufficient condition for $A$ to be isomorphic to an
algebra of global sections of a sheaf with fibers $A_i = A/\theta_i$,
for $i \in I$ is given below:
\begin{definition}
A family $(c_i)_{i \in I}$ of elements of $A$ is
said to be {\em global}\index{global family of elements} 
with respect to $(\theta_i)_{i \in I}$ if for every
$i \in I$ there exist $a_1^i,\dots,a_n^i, b_1^i,\dots,b_n^i \in A$
such that: 
\begin{enumerate}
\item[(i)] $(a^i_j, b^i_j) \in \theta_i$ for every $j = 1,\dots,n$, 

\item[(ii)] If $(a^i_j, b^i_j) \in \theta_k$ for every $j =
1,\dots,n$ then $(c_k, c_i) \in \theta_k$. 
\end{enumerate}
\end{definition}
\begin{theorem}[\cite{Davey}] Let $(\theta_i)_{i \in I}$ be a family of
congruences on an algebra $A$ such that $A$ is a subdirect product
of $(A/\theta_i)_{i \in I}$.  Endow $I$ with its coarsest
S-topology. Then $\alpha : A \rightarrow \Gamma(I, F_A)$ is an
isomorphism iff for every family of elements $(c_i)_{i \in
I}$ global with respect to $(\theta_i)_{i \in I}$, there is a $c \in
A$ with $(c, c_i) \in \theta_i$ for every $i \in I$.  
\end{theorem}

\section{Appendix. Partially commutative monoids}
\label{app:pcm}

If $G = (A, D)$ is a dependency graph, we denote by $M(G)$ the 
quotient $A^*/\theta$, where $\theta$ is the congruence 
generated by $\{ (a_1 a_2, a_2 a_1) \mid (a_1, a_2) \not\in D \}$ (a 
free partially commutative monoid). 

\begin{theorem}[Corollary 1.4.5~ in \cite{Diekert90}] 
Let $G$ be an undirected graph
and $\{ G_j  \mid  j \in J \}$ be a {\em finite} family of subgraphs of $G$. 
For $j \in J$ let $\pi_j : M(G) \rightarrow M(G_j)$ be the canonical 
projection  and $\pi : M(G) \rightarrow \prod_{j \in J} M(G_j)$ be the 
homomorphism into the direct product defined by $\pi(t) = (\pi_j(t))_{j \in J}$.
Then $\pi$ is injective iff $G {=} {\bigcup}_{j \in J} G_j$.
\label{GET}
\end{theorem}
\noindent If $\{ M_j  \mid  j \in J \}$ is a family of non-trivial 
free partially commutative monoids then  $ \prod_{j \in J} M_j$ 
is free partially 
commutative iff $J$ is finite \cite{Diekert90}.
If $\{ G_j  \mid  j \in J \}$ is not finite, then -- assuming that 
for every vertex $x$ of $G$ there are finitely many $j \in J$ 
such that $x$ is a vertex of $G_j$ -- 
there is an injective morphism 
$M(G) \hookrightarrow \bigoplus_{j \in J} M(G_j)$, 
where $\bigoplus_{j \in J} M(G_j) = \{ (m_j)_{j \in J}  
\mid  m_j \in M(G_j) \mbox{ for all } j \in J, m_j = \varepsilon 
\mbox{ a.e.}\footnote{a.e. means {\em almost everywhere}} \}$
\cite{Diekert90}, p.27.


\begin{thebibliography}{15}


\bibitem[1]{ChangKeisler90}
C.C. Chang and H.J. Keisler.
\newblock {\em Model Theory}.
\newblock North-Holland, Amsterdam, 3rd edition, 1990.


\bibitem[2]{Davey}
B.A.~Davey.
\newblock {Sheaf spaces and sheaves of universal algebras}.
\newblock Math.\ Zeitschrift, 134:275--290, 1973.


\bibitem[3]{Diekert90}
V.~Diekert.
\newblock {Combinatorics on Traces}.
\newblock In {\em LNCS 454}. Springer Verlag, 1990.


\bibitem[4]{Fajstrup06}
L.~Fajstrup, M.~Raussen, E.~Goubault.
\newblock Algebraic topology and concurrency, I. 
\newblock {\em Theoretical Computer Science} 357, pages 241--278, 2006.


\bibitem[5]{Gaucher03}
P.~Gaucher, {\'E}.~Goubault. 
\newblock Topological Deformation of Higher Dimensional Automata.
\newblock {\em Homology, Homotopy, Appl.}, 5(2):39--82, 2003.


\bibitem[6]{Goguen92}
J.A. Goguen.
\newblock Sheaf semantics for concurrent interacting objects.
\newblock {\em Mathematical Structures in Computer Science}, 11:159--191, 1992.


\bibitem[7]{Herlihy95}
M.~Herlihy, N.~Shavit..
\newblock The topological structure of asynchronous computation.
\newblock {\em Journal of the ACM}, 46: 858-923, 1999.


\bibitem[8]{Jacobs-Sofronie-pdpar06}
S. Jacobs and V. Sofronie-Stokkermans.
\newblock Applications of hierarchical reasoning 
in the verification of complex systems. 
\newblock Electronic Notes in Computer Science 174(8), pages 39-54, 2007.
(Selection of the papers presented at 
the IJCAR'06 workshop Pragmatics of Decision Procedures in Automated Reasoning
(PDPAR'06).)


\bibitem[9]{Johnstone82}
P.~Johnstone.
\newblock {\em Stone Spaces}.
\newblock Cambridge Studies in Advanced Mathematics 3. Cambridge University
  Press, 1982.


\bibitem[10]{Krauss-Clark-1979}
P.H. Krauss and D.M. Clark.
\newblock Global subdirect products.
\newblock {\em Memoirs of the AMS}, 17(210):1--109, 1979.


\bibitem[11]{Kroger87}
F.~Kr{\"o}ger.
\newblock {\em Temporal Logic of Programs}, volume~8 of {\em EATCS Monographs
  on Theoretical Computer Science}.
\newblock Springer Verlag, 1987.


\bibitem[12]{MacLaneMoerdijk92}
S.~Mac~Lane and I.~Moerdijk.
\newblock {\em Sheaves in Geometry and Logic}.
\newblock Universitext. Springer Verlag, 1992.


\bibitem[13]{MonteiroPereira86}
L.~Monteiro and F.~Pereira.
\newblock A sheaf theoretic model for concurrency.
\newblock {\em Proc. Logic in Computer Science (LICS'86)}, 1986.


\bibitem[14]{Pratt1}
V.~Pratt.
\newblock Modeling concurrency with geometry.
\newblock {\em Proc. 18th Symposium on Principles of Programming Languages} 
pages 311-322, ACM Press New York USA, 1991.


\bibitem[15]{Pratt2}
V.~Pratt.
\newblock Higher-dimensional automata revisited.
\newblock {\em Mathematical Structures in Computer Science}, 10(4), 2000.


\bibitem[16]{Sofronie96a}
V.~Sofronie.
\newblock Towards a sheaf theoretic approach to cooperating agents scenarios.
\newblock In J.~Calmet, J.A. Campbell, and J.~Pfalzgraf, editors, {\em
  Proc.\ of the International Conference Artificial Intelligence and Symbolic Mathematical Computation (AISMC-3)}, LNCS 1138, pages 289--304.
  Springer Verlag, 1996.


\bibitem[17]{Sofronie97}
V.~Sofronie-Stokkermans.
\newblock {\em Fibered Structures and Applications to Automated Theorem Proving
  in Certain Classes of Finitely-Valued Logics and to Modeling Interacting
  Systems}.
\newblock PhD thesis, RISC-Linz, J.\ Kepler University Linz, 1997.


\bibitem[18]{Sofronie-Stokkermans99}
V.~Sofronie-Stokkermans and K.\ Stokkermans.
\newblock {\em Modeling Interaction by Sheaves and Geometric Logic}.
\newblock In G. Ciobanu and Gh. Paun eds, Proc.\ International Conference 
Fundamentals of Computation Theory (FCT'99), LNCS 1684, pages 512-523, Springer Verlag, 1999.

\end{thebibliography}
\end{document}